\documentclass[]{article}
\pdfoutput=1
\usepackage{amsmath}
\usepackage{amssymb}
\usepackage[pdftex]{graphicx}
\usepackage{mathptmx}
\usepackage{float}
\usepackage{caption}
\usepackage{exscale,relsize}
\usepackage[mathscr]{eucal}
\usepackage[pdfstartview={XYZ null null 1.10},pagebackref]{hyperref}

\DeclareCaptionLabelFormat{New}{Fig.\nobreakspace#2}
\usepackage{tocloft}

\newcommand{\asin}{\mathrm{asin}}

\newcommand{\atan}{\mathrm{atan}}
\newcommand{\catan}{\mathrm{catan}}

\newcommand{\xdddot}{\begin{picture}(5.7,0)\put(1,6.2){\circle*{1}}\put(3,6.2){\circle*{1}}\put(5,6.2){\circle*{1}}\put(1,0){$x$}\end{picture}}
\newcommand{\pdddot}{\begin{picture}(5.7,0)\put(.9,6.2){\circle*{1}}\put(2.9,6.2){\circle*{1}}\put(4.9,6.2){\circle*{1}}\put(-1.0,0){$p$}\end{picture}}

\newcommand{\diffd}{d}
\newcommand{\diffdspc}{d\,}

\newcommand{\imgi}{i}

\begin{document}

\title{Wave, Particle, or a Third Possibility?}
\author{John T. Brooker \\ e-mail: jtbarchive@hotmail.com}

\maketitle

\noindent Keywords: DEO, discrete extension, discretely extended, dynamic collapse, dynamic quantization, operator-free, primitive ontology, quasi-discrete, quasi-Newtonian

\bigskip

\begin{abstract}

\indent This paper presents an alternative quantum theory, the \emph{Theory of Discrete Extension}, which avoids many of the conceptual problems of standard quantum mechanics.  It is a deterministic, dynamic collapse theory with a well-defined primitive ontology.

\indent In place of the dual, classical concepts of wave and particle, the unitary, non-classical concept of a \emph{discretely extended object} emerges directly from the theory's dynamic equations as a primitive ontology.  Because this ontology is unitary, the theory avoids the dilemmas of wave-particle dualism and complementarity.

\indent Furthermore, the theory's dynamic equations generate correct, quasi-discrete values of action increments and energy levels without recourse to the operator formalism and eigenvalue postulate of standard quantum mechanics.  Quantization of the harmonic oscillator provides a simple illustration.

\indent The theory provides insight into the nature of a number of quantum effects such as the zero-point energy of the harmonic oscillator.  It also makes a number of predictions that distinguish it from standard quantum mechanics and from Bohmian mechanics.

\end{abstract}

\newpage

\tableofcontents

\newpage

\section{Introduction}

\subsection{Intent and Scope}

In a course on quantum mechanics, students are confronted, at the outset, with a number of enigmatic fundamentals such as wave-particle dualism, superposition of states, wave function collapse, indeterminism, and the substitution of mathematical operators for real-valued observables.  These elements lead to unresolved questions of interpretation of the theory.  After sufficient exposure, a student may come to accommodate these unfamiliar tenets, but unease with such concepts is likely to persist, nevertheless. These issues can be troubling, not only for many students, but for professional physicists at the highest level.  Steven Weinberg, for example, has expressed the view that ``...today there is no interpretation of quantum mechanics that does not have serious flaws...we ought to take seriously the possibility of finding some more satisfactory other theory...''~\cite{sW-2015}. 

In these unsettled circumstances, this paper offers an alternative quantum theory, the \emph{Theory of Discrete Extension}, which avoids many of the conceptual problems of standard quantum mechanics.  It is based on a classically familiar mathematical formalism and has an unambiguous primitive ontology, the \emph{discretely extended object} (DEO).  It also includes a deterministic, dynamic collapse process by which it avoids the measurement problem of standard quantum mechanics.  Outside observers who make measurements on quantum systems play no fundamental role in the theory.

A body of results has been obtained concerning the development, interpretation, and application of the theory and will be presented in this and subsequent papers.  The principal goals of this initial paper are limited, however, to three.  The first is to introduce the concept of a DEO (section~\ref{sec:DEO definition}) and to propose it as a characterization of the nature of quantum-level entities, \emph{i.e.}, as a primitive ontology.  The paper offers this non-classical concept as a unitary alternative to the dual, classical concepts of wave and particle.  Because its ontology is unitary, the theory avoids the dilemmas of wave-particle dualism and complementarity.

The second goal is to develop the mathematical theory of DEO dynamics (section~\ref{sec:Mathematical Theory}), to then show that the concept of a DEO arises directly from the dynamic equations of the theory, and, finally, to illustrate examples of DEO dynamics for simple systems.  The paper derives the theory's dynamic equations from the unmodified, linear Schr\"odinger equation.  The result is a deterministic extension of Newtonian mechanics that takes into account the nonzero value of Planck's constant, $h$.  The paper proposes that this theory is valid for systems in the quantum domain.

The third goal is to show that correct, quasi-discrete action increments and energy levels for bound-state systems are generated by the theory's dynamic equations, alone, without recourse to the operator formalism and eigenvalue postulate of standard quantum mechanics.  In his first paper on wave mechanics Schr\"odinger notes that he has traced the origin of integral quantum numbers to the ``finiteness and single-valuedness'' of the wave function and that this insight is an advance over the ad hoc postulate of whole numbers in the old quantum theory~\cite{eS-26}.  The theory of discrete extension provides a further advance by tracing quantization directly to its dynamic equations, alone.

The paper applies this quantization procedure to the harmonic oscillator (section~\ref{sec:Oscillator}) and shows the results in figure~\ref{osc full-cycle action vs energy high activ}.  This simple exercise is sufficient to show how the theory differs from standard quantum mechanics in physical, mathematical, interpretive, and predictive content.  In particular, it shows how quantization can be achieved within a mathematical structure similar to that of Newtonian mechanics.

In its current form the theory deals with concepts and systems within the realm of elementary quantum mechanics only.  This limited domain seems to be the appropriate one, however, when the fundamentals of quantum theory are being examined.  Not only are the main conceptual problems already present at that level, but the elementary theory also serves as the foundation for more advanced theories.

The principles of the theory of discrete extension have not yet been applied to relativistic systems, and its implications for quantum field theory have yet to be considered.  Nevertheless, the theory already offers insight into the nature of a number of quantum effects such as the zero-point energy of the harmonic oscillator (section~\ref{sec:Zero-Point Energy}).  It also makes a number of predictions that distinguish it from standard quantum mechanics and from Bohmian mechanics.  For example, it predicts that the natural frequency of the harmonic oscillator depends on the oscillator's energy (figure~\ref{quasi-Newt osc freq vs energy}).  It may be possible to look for this effect experimentally in the field of nanotechnology.
\subsection{Additional Results}
The following additional results have been obtained and will be presented in subsequent papers.  Application of the theory to the double slit system shows that the associated interference pattern is the cumulative result of a series of random, but deterministic, events.  An analysis of two coupled oscillators demonstrates the quasi-discrete transfer of energy between subsystems.  An analysis of the hydrogen atom shows that the theory generates correct, quasi-discrete values of quantum numbers for a system with three degrees of freedom.  An analysis of distant, entangled subsystems demonstrates the non-local influence of one subsystem on another and introduces the concept of \emph{energy of entanglement}.  Finally, an interaction involving a DEO is shown to cause a dynamic collapse in which the DEO evolves continuously to a collapsed state according to the dynamic equations of the theory.
\subsection{Discrete Extension and the Concept of a DEO} \label{sec:DEO definition}
In the 20th century a number of previously unknown physical effects were discovered such as the Compton effect and electron diffraction.  These discoveries revealed that light and matter have both particle-like and wave-like aspects to their behaviors.  Over the years, various attempts have been made to interpret these new phenomena in terms of classical wave and particle concepts that had been carried over from the 19th century.

The most widely accepted interpretation, the Copenhagen interpretation, incorporates these classical concepts through the notions of wave-particle dualism and complementarity~\cite{dB-51}.  Other interpretations, such as Schr\"odinger's unitary wave theory~\cite{eS-27}, attempt to explain particle-like behavior on the basis of continuum dynamics.  Conversely, still other interpretations, such as Land\'e's unitary particle theory~\cite{aL-60}, attempt to explain wave-like interference phenomena on the basis of particle dynamics.  Finally, interpretations such as the de~Broglie--Bohm pilot wave theory~\cite{dB-52-1} postulate a dual physical basis of particles together with an objectively real field.

These interpretations all rely on the classical wave and particle concepts.  In contrast, the theory of discrete extension is based on the non-classical, unitary concept of a DEO.

A classical particle is a point-like entity and can be described, therefore, as an \emph{unextended object}.  A classical wave or field is a space-filling entity and can be described as a \emph{continuously extended object}.  In contrast to these classical concepts, a DEO is a single, indivisible entity with multiple, spatially isolated, point-like manifestations.  Therefore, with similar terminology, it can be described as a \emph{discretely extended object}.  In special cases a DEO can be unextended like a particle, as shown below (section~\ref{sec:free DEO dynamics}).

In standard quantum mechanics the notion of a particle trajectory is rejected in accordance with the uncertainty principle.  A DEO, on the other hand, has a well-defined spatial trajectory and a continuous space-time world-line.  Despite its partitioned appearance, a DEO is an undivided entity whose spatially isolated manifestations, called DEO points, have no individual existence apart from the DEO as a whole.  An entity will be considered to be undivided or whole if its world-line is continuous.

At first, the characteristics of having a partitioned appearance and being undivided seem to be in conflict.  There is no inconsistency, however, because, as shown below (e.g. equation~\eqref{free DEO world line}), DEO world-lines are, in general, oscillatory and will therefore have multiple intersections with lines of simultaneity. 

Being both point-like and extended, a DEO is a hybrid object that combines certain attributes of both particles and waves.  In particular, the point-like manifestations of a free DEO can have a spatial distribution that is quasi-periodic.  Therefore, in addition to its point-like aspect, such a DEO possesses an attribute that is analogous to the wavelength of a harmonic wave.  In future papers it will be shown that the dynamics of this hybrid object can lead to phenomena, as in the double slit experiment, that are normally attributed to the interference of physical waves.

Although it is possible for a DEO to exist in an unextended state, no such DEOs are likely to be found in nature.  Any DEO that is found in, or has been specially prepared in, an unextended state is nearly certain to become discretely extended upon interaction with the environment.  This behavior will be illustrated below (section~\ref{sec:barrier interaction}) by an encounter between an unextended free DEO and a rectangular barrier.  Thus, naturally occurring DEOs are almost certain to be discretely extended due to a history of countless interactions with the environment.

\section{The Mathematical Theory of DEO Dynamics} \label{sec:Mathematical Theory}
\subsection{Mathematical Framework}
The mathematical theory of DEO dynamics is an extension of Newtonian mechanics that takes into account the non-zero value of Planck's constant, $h$.  As such, it is proposed here that the theory is valid for systems in the quantum domain.  The starting point for this extension is the Hamilton--Jacobi formulation of Newtonian theory, a choice that is motivated by the following considerations.
\subsubsection{The Polar Decomposition of Schr\"odinger's Equation}
In standard quantum mechanics the wave function, $\Psi$, for a one-particle system satisfies Schr\"odinger's equation,
\begin{equation} \label{Sch eqn}
\dfrac{\hslash^{2}}{2m}\nabla^{2}\Psi-V(\vec{x},t)\Psi+\imgi\,\hslash\,\dfrac{\partial{\Psi}}{\partial{t}} = 0\,,
\end{equation}
where $m$ is the particle's mass, $V(\vec{x},t)$ is the potential function, and $\hslash = h/2\pi$.  The complex wave function can be expressed in the polar form $\Psi = R\,\exp(\imgi\,S/\hslash)$, and Schr\"odinger's equation is then equivalent to the following coupled, real equations for the amplitude function, $R$, and the phase function, $S$\,:
\begin{equation}
\nabla{\cdot\!\left(\!R^{2}\dfrac{\nabla{S}}{m}\right)}+\dfrac{\partial}{\partial{t}}\left(R^{2}\right) = 0
\label{continuity eqn}
\end{equation}
\begin{equation}
\dfrac{\left(\nabla{S}\right)^{2}}{2m}+V(\vec{x},t)+\dfrac{\partial{S}}{\partial{t}}-\dfrac{\hslash^{2}}{2m}\dfrac{\nabla^{2}{R}}{R} = 0\,.
\label{qm H-J eqn}
\end{equation}
This decomposition of Schr\"odinger's equation is used in the development of both the de~Broglie--Bohm pilot wave theory~\cite{dB-52-1} and the theory of discrete extension introduced here.

Equation~\eqref{continuity eqn} is a continuity equation that, in standard quantum mechanics, expresses the conservation of probability.  The real quantities
\begin{equation*}
R^{2}\dfrac{\nabla{S}}{m} = \dfrac{\imgi\,\hslash}{2m}\left(\Psi\,\nabla{\Psi^{*}}-\,\Psi^{*}\,\nabla{\Psi}\right) = \vec{j} \qquad\qquad \text{and} \qquad\qquad R^{2} = \Psi^{*}\,\Psi = \rho
\end{equation*}
are interpreted there as a probability current density and a probability density, respectively.  $\Psi^{*}$ is the complex conjugate of $\Psi$.

\subsubsection{The Generalization of Hamilton's Principal Function}

In the classical limit with $\hslash = 0$, equation~\eqref{qm H-J eqn} is uncoupled from $R$ and reduces to the time-dependent Hamilton--Jacobi equation of Newtonian theory.  Any equation that is uncoupled from $R$ will subsequently be called a direct equation.  Thus, for $\hslash = 0$, equation~\eqref{qm H-J eqn} reduces to a direct equation for Hamilton's principal function, $S_{N}^{}$, where the subscript, $N$, indicates a quantity from Newtonian theory.

For stationary states of a conservative system with one degree of freedom, equations~\eqref{continuity eqn} and~\eqref{qm H-J eqn} can be uncoupled even when $\hslash \neq 0$.  In that case the wave function has the form $\Psi = R(x)\exp[\,\imgi\,(W\!(x)-E\,t)/\hslash\,]$ where the constant $E$ is the energy of the system.  Then, $\partial{R}/\partial{t} = 0$ and $\partial{S}/\partial{t} = -E$, and the function $W(x)$ satisfies the following direct, third-order equation:
\begin{equation}
\left(\dfrac{\diffd W}{\diffd x}\right)^{\!\!2}\!\left[\left(\dfrac{\diffd W}{\diffd x}\right)^{\!\!2}\!-2m[\,E-V(x)\,]\right]+\left(\dfrac{\hslash}{2}\right)^{\!\!2}\!\left[2\,\dfrac{\diffd W}{\diffd x}\,\dfrac{\diffd^{3}W}{\diffd x^{3}}-3\!\left(\dfrac{\diffd^{2}W}{\diffd x^{2}}\right)^{\!\!2}\right] = 0\,.
\label{direct eqn for W}
\end{equation}
For $\hslash = 0$ this equation reduces to the time-independent Hamilton--Jacobi equation of Newtonian theory.

The fact that equations~\eqref{qm H-J eqn} and~\eqref{direct eqn for W} from standard quantum mechanics reduce to Newtonian Hamilton--Jacobi equations for $\hslash = 0$ suggests that Newtonian Hamilton--Jacobi theory can be broadened into a quantum theory when $\hslash \neq 0$.  Therefore, the theory of discrete extension adopts as its framework the entire Hamilton--Jacobi formalism of Newtonian theory but with the Schr\"odinger phase function, $S$, playing the role of Hamilton's principal function.  The theory adopts this formalism not only for the one-particle systems discussed above, but for systems of arbitrary complexity.  In all cases the generalized principal function, $S$, will satisfy a direct, higher-order equation determined, implicitly, by Schr\"odinger's equation.

\subsubsection{The Generalized Hamilton--Jacobi Equation} \label{sec:gen H-J eqn}

At first this prescription seems impossible.  On the one hand, Hamilton--Jacobi theory is based on the mathematical relation that exists between any first-order partial differential equation and its associated system of characteristic, first-order, ordinary differential equations (Hamilton's canonical equations in the case of Newtonian theory).  On the other hand, no such relation exists for higher-order partial differential equations such as the implicit, direct equation for $S$.

This higher-order equation, however, may have an implicit, first-order, intermediary integral that shares solutions with its parent equation.  In the theory of discrete extension an intermediary integral of this sort plays the role of a generalized Hamilton--Jacobi equation.  Being of first-order, it is mathematically qualified for that role.

Let $\mathscr{N}$ be the number of degrees of freedom of a given system, and let \\ $n = 1,2,\ldots,\mathscr{N}$ be an index for the coordinates, $q_n^{}$, in the configuration space of that system.  Also, let $\nu = 0,1,2,\ldots,\mathscr{N}$ be an index in the extended configuration space that includes time as the coordinate $q_0^{} = t$.

Reduction of the higher-order equation for $S$ to a first-order, intermediary integral will then result in a generalized Hamilton--Jacobi equation of the form
\begin{equation}
K(q_\nu^{},\dfrac{\partial S}{\partial q_\nu^{}},\gamma) = 0
\label{gen H-J eqn}
\end{equation}
where $\gamma$ is a set of arbitrary integration constants that are introduced in the order-reduction process.  The function $K(q_\nu^{},p_\nu^{},\gamma)$, which is identically zero, serves as an extended Hamiltonian function for the system.  The quantities $(q_\nu^{},p_\nu^{})$ in this Hamiltonian are canonical pairs of coordinates and momenta in the extended phase space of the system.  They include the canonical pair $(q_0^{},p_0^{}) = (t,-E)$ where $E = -\partial S / \partial t$ is the energy of the system (not necessarily constant).

The Hamilton--Jacobi formalism requires a complete solution of equation~\eqref{gen H-J eqn}.  In addition to an additive constant that can be ignored, such a solution, $S(q_\nu^{},\alpha_n^{},\gamma)$, will contain $\mathscr{N}$ essential Hamilton--Jacobi constants, $\alpha_n^{}$, as well as the set, $\gamma$, of order-reduction constants.  It therefore involves two distinct sets of arbitrary constants, different from each other in both origin and mathematical status.

Since $S$ depends on the arbitrary constants in the set $\gamma$, the Hamilton--Jacobi motion equations, $\partial S/\partial \alpha_n^{} = \beta_n^{}$, will also depend on them.  These constants comprise non-Newtonian data that must be specified in addition to the usual Newtonian initial conditions in order to uniquely define the behavior of the system.  The theory of discrete extension is, therefore, richer than Newtonian theory in that it permits infinite families of diverse DEO dynamics for any given set of Newtonian initial conditions.

\subsubsection{The Roles of Schr\"odinger's Equation and the Wave Function}

In the theory of discrete extension Schr\"odinger's equation~\eqref{Sch eqn} is viewed only as an auxiliary mathematical tool for obtaining the real-valued function, $S$, while avoiding the inherent difficulties in the derivation and solution of a non-linear, direct equation for $S$.  The following progression summarizes the logical steps that relate Schr\"odinger's equation to the generalized Hamilton--Jacobi equation~\eqref{gen H-J eqn}:
\begin{enumerate}
  \item Schr\"odinger's equation for the wave function, $\Psi = R\,\exp(\imgi\,S/\hslash)$, is second-order, linear, and complex.  It can be decomposed into
	\item Two coupled equations for $R$ and $S$ which are second-order, non-linear, and real.  Elimination of $R$ between them leads implicitly to
	\item A direct equation for $S$ which is higher-order, non-linear, and real.  An implicit, first-order, intermediary integral of this implicit equation is then
	\item The generalized Hamilton--Jacobi equation, $K(q_\nu^{},\partial S/\partial q_\nu^{},\gamma) = 0$, for $S$.
\end{enumerate}

In the theory of discrete extension the wave function, $\Psi$, has neither the status of a physical field nor an interpretation as either a probability amplitude or a state vector.  Its role is simply to lend its phase function, $S$, for use as a generalized principal function in the Hamilton--Jacobi formalism.

\subsubsection{Action Functions and Phase Functions}

In Newtonian theory the solution, $S_{N}^{}$, of the Hamilton--Jacobi equation is closely related to the Newtonian action function.  Therefore, in the theory of discrete extension, Schr\"odinger's phase function, $S$, which satisfies a generalized Hamilton--Jacobi equation, will be viewed as a generalized action function and will be referred to simply as the action.  Recognition of a close connection between action functions and phase functions is, of course, not new.  In the path integral formulation of standard quantum mechanics, for example, $1 / \hslash$ times the Newtonian action integral, $\int\!L\,\diffd t$, over a given classical path is adopted as the phase of the contribution that the path makes to the total quantum amplitude.  Thus, in that theory, the conceptual order is from an action function to a phase function.  In the theory of discrete extension, however, the order is reversed: $\hslash$ times the phase of the Schr\"odinger wave function is used in the Hamilton--Jacobi formalism as a non-classical action function for a theory of DEO dynamics.
\subsection{Contrast with the de~Broglie--Bohm Pilot Wave Theory}

The development of the de~Broglie--Bohm pilot wave theory and the theory of discrete extension both begin with the Hamilton--Jacobi formulation of Newtonian mechanics, and both make use of the phase function, $S$, in the polar form of Schr\"odinger's wave function.  The two theories, however, have many differences.  Two important areas of disagreement are quantization procedures and motion equations.
\subsubsection{Quantization Procedures}

The de~Broglie--Bohm theory made an important conceptual advance by demonstrating that neither indeterminism nor complementarity are necessary parts of a coherent interpretation of quantum theory.  The theory can be criticized, however, for interpretive problems of its own.  It postulates a dual physical basis consisting of point particles together with an ``objectively real field''~\cite{dB-52-2,jB-93}, \emph{i.e.} a pilot wave, that guides the motions of the particles.  Insisting that the pilot wave is objectively real provides the theory with a \emph{physical} justification for requiring that the mathematical representation of this field be finite, continuous, and single-valued.  That is, it justifies the use of eigenvalues to achieve quantization. In this way, postulating the reality of the pilot wave ensures that the theory will generate the same quantization results as standard quantum mechanics~\cite{dB-52-3}.

There are strong arguments, however, against the plausibility of this postulate.  Unlike a physical object such as the electromagnetic field, the pilot field is sourceless and is without an energy density.  Furthermore, the influence of the pilot field is independent of its magnitude; it can be multiplied by an arbitrary constant without changing its effect on the motion of the particles.  Finally, for a system of $N$ particles, it is represented mathematically as a function on the $3N$-dimensional configuration space of the particles rather than in physical space.  For these reasons the pilot field should be viewed as nothing more than an auxiliary mathematical construct.  As a result, there is no \emph{physical} basis for requiring it to be finite, continuous, and single-valued.  The de~Broglie--Bohm theory, therefore, lacks a physical justification for using eigenvalues to achieve quantization.  (This lack of physical justification is true of standard quantum mechanics, as well.)

The theory of discrete extension, on the other hand, leads to the unitary primitive ontology of a DEO and does not postulate the existence of an associated physical field.  Therefore, in this theory, the mathematical restrictions stated above for the pilot wave are irrelevant.  As shown below (section~\ref{sec:Quantization}), quantization emerges directly from the dynamic equations of the theory without the need for these supplementary conditions.


\subsubsection{Motion Equations}

The de~Broglie--Bohm theory makes use of only half of the Hamilton--Jacobi formalism while disregarding the other half.  It postulates that the momentum components of a particle are given by $p_{n}^{} = \partial S/\partial q_{n}^{}$ where the function $S$ is adopted as a generalization of Hamilton's principal function.  It also postulates, however, that the velocity components of the particle are given by the Newtonian expression $\textsl{v}_{n}^{} = p_{n}^{}/m$ where $m$ is the particle's mass.  The purpose of this second postulate is to preserve the standard probability interpretation of equation~\eqref{continuity eqn} for an ensemble of particles~\cite{dB-52-4}.  Taken together, these two postulates produce motions that are in conflict with the Hamilton--Jacobi motion equations, $\partial S/\partial \alpha_{n}^{} = \beta_{n}^{}$, where $\alpha_{n}^{}$ and $\beta_{n}^{}$ are the usual Hamilton--Jacobi constants.  The de~Broglie--Bohm theory ignores this set of motion equations.

The theory of discrete extension, on the other hand, employs the entire Hamilton--Jacobi formalism, including the motion equations.  In doing so, it leads to a new relation between momentum and velocity and a new expression for the probability density of position.  Furthermore, the concept of a DEO arises naturally from this mathematical structure due to the oscillatory nature of the world-lines produced by these motion equations.

\subsection{Action, Momentum, and World-Line Functions in Terms of the Wave Function}

This section derives the action, momentum, and world-line functions for a given system in terms of a particular wave function, $\Psi$, for that system.  It will be seen in later sections that this particular wave function consists of a specific linear combination of linearly independent solutions of Schr\"odinger's equation and that the coefficients of that linear combination are defined in terms of the non-Newtonian constants that comprise the set $\gamma$ introduced in section~\ref{sec:gen H-J eqn}.
\subsubsection{The Action Function}
For now, assume that the required linear combination for $\Psi$ is given, and let its real and imaginary parts be $\Psi\!R$ and $\Psi\!I$, respectively.  The action function, $S(q_{\nu}^{},\alpha_{n}^{},\gamma)$, can then be derived in terms of the wave function, $\Psi = \Psi\!R+\imgi\,\Psi\!I = R\,\exp(\imgi\,S/\hslash)$, and its value, $\Psi_{0}^{} = \Psi\!R_{0}^{}+\imgi\,\Psi\!I_{0}^{} = R_{0}^{}\,\exp(\imgi\,S_{0}^{}/\hslash)$, at a selected reference point, $q_{\nu}^{} = (q_{\nu}^{})_{0}^{}$, in the extended configuration space of the given system.  Let $S_{0}^{} = 0$ by convention.  The wave function then has the form
\begin{equation} \label{Wave Function}
\Psi = \Psi_{0}^{}\left(\dfrac{R}{R_{0}^{}}\right)\exp\!\left(\dfrac{\imgi\,S}{\hslash}\right)\,.
\end{equation}
Solving for $S$ gives
\begin{equation*}
S = \hslash\;\atan\!\left(\dfrac{\Psi\!R_{0}^{}\;\Psi\!I-\Psi\!I_{0}^{}\;\Psi\!R}{\Psi\!R_{0}^{}\;\Psi\!R+\Psi\!I_{0}^{}\;\Psi\!I}\right)\,.
\end{equation*}
This equation is independent of $R$ and is therefore a direct equation for $S$.  Applying the arctangent addition theorem leads to the simplified equation
\begin{equation} \label{action}
S = \hslash\left[\atan\!\left(\dfrac{\Psi\!I}{\Psi\!R}\right)-\atan\!\left(\dfrac{\Psi\!I_{0}^{}}{\Psi\!R_{0}^{}}\right)\right]\,.
\end{equation}

For a stationary state of a conservative system with energy $E$, the wave function has the form $\Psi = \psi\,\exp(-\imgi\,E\,t/\hslash)$ where $\psi = \psi R+\imgi\,\psi I = R\,\exp(\imgi\,W\!/\hslash)$ is the time-independent wave function.  The action function is then $S = W-Et$.  Substituting these expressions for $\Psi$ and $S$ into equation~\eqref{Wave Function} gives
\begin{equation*}
\psi = \psi_{0}^{}\left(\dfrac{R}{R_{0}^{}}\right)\exp\!\left(\dfrac{\imgi\,W}{\hslash}\right)
\end{equation*}
which is of the same form as equation~\eqref{Wave Function}.  Therefore, solving for the abbreviated action, $W$, gives
\begin{equation} \label{abbreviated action}
W = \hslash\left[\atan\!\left(\dfrac{\psi I}{\psi R}\right)-\atan\!\left(\dfrac{\psi I_{0}^{}}{\psi R_{0}^{}}\right)\right]\,
\end{equation}   
which is identical to equation~\eqref{action} except that $\Psi$ has been replaced by $\psi$.
\subsubsection{Generalized Momentum Components}
In the theory of discrete extension, as in Newtonian theory, the derivative of $S$ with respect to a generalized coordinate, $q_{\nu}^{}$, is the generalized momentum, $p_{\nu}^{}$, conjugate to $q_{\nu}^{}$.  In terms of the real and imaginary parts of the wave function, the momentum components, $p_{\nu}^{}$, are given by
\begin{equation*}
p_{\nu}^{} = \dfrac{\partial{S}}{\partial{q_{\nu}^{}}} = \dfrac{\hslash\left(\!\Psi\!R\,\dfrac{\partial{\Psi\!I}}{\partial{q_{\nu}^{}}}-\Psi\!I\,\dfrac{\partial{\Psi\!R}}{\partial{q_{\nu}^{}}}\right)}{\Psi\!R^{\,2}+\Psi\!I^{\,2}}\,.
\end{equation*}
This expression is independent of the reference point.

In standard quantum mechanics the probability density for position and the probability current density in the $q_{n}^{}$ direction are
\begin{equation*}
\rho = \Psi R^{\,2}+\Psi I^{\,2} \qquad \text{and} \qquad j_{n}^{} = \dfrac{\hslash}{m}\left(\!\Psi\!R\,\dfrac{\partial{\Psi\!I}}{\partial{q_{n}^{}}}-\Psi\!I\,\dfrac{\partial{\Psi\!R}}{\partial{q_{n}^{}}}\right)\;,
\end{equation*}  
respectively.  Furthermore, if $\hslash$ is allowed to approach zero in the standard theory, the resultant classical limit leads to an interpretation in which $\Psi$ describes an ensemble of particles with momentum components $p_{n}^{} = m\,j_{n}^{}/\rho$~\cite{aM-66}.  In the theory of discrete extension the probability interpretations of $\rho$ and $j_{n}^{}$ do not apply, but it can be seen that, for each DEO point, the equation $p_{n}^{} = m\,j_{n}^{}/\rho$ is true even when $\hslash \neq 0$.

For a stationary state of a conservative system $S = W-Et$, and the momentum components are
\begin{equation*}
p_{n}^{} = \dfrac{\partial{S}}{\partial{q_{n}^{}}} = \dfrac{\partial{W}}{\partial{q_{n}^{}}} =  \dfrac{\hslash\left(\!\psi\!R\,\dfrac{\partial{\psi\!I}}{\partial{q_{n}^{}}}-\psi\!I\,\dfrac{\partial{\psi\!R}}{\partial{q_{n}^{}}}\right)}{\psi\!R^{\,2}+\psi\!I^{\,2}}\,.
\end{equation*}

\subsubsection{Motion Equations}

In the theory of discrete extension, as in Newtonian theory, the motion equations are obtained from the derivatives of $S$ with respect to the parameters, $\alpha_{n}^{}$, in a complete solution, $S(q_{\nu}^{},(q_{\nu}^{})_{0}^{},\alpha_{n}^{},\gamma)$, of the (generalized) Hamilton--Jacobi equation.

The derivative of $S$ with respect to the parameter $\alpha_{n}^{}$ is
\begin{equation} \label{dS dalpha}
\dfrac{\partial{S}}{\partial{\alpha_{n}^{}}} = \dfrac{\hslash\left(\!\Psi\!R\,\dfrac{\partial{\Psi\!I}}{\partial{\alpha_{n}^{}}}-\Psi\!I\,\dfrac{\partial{\Psi\!R}}{\partial{\alpha_{n}^{}}}\right)}{\Psi\!R^{\,2}+\Psi\!I^{2}}-\dfrac{\hslash\left(\!\Psi\!R_{0}^{}\,\dfrac{\partial{\Psi\!I_{0}^{}}}{\partial{\alpha_{n}^{}}}-\Psi\!I_{0}^{}\,\dfrac{\partial{\Psi\!R_{0}^{}}}{\partial{\alpha_{n}^{}}}\right)}{\Psi\!R_{0}^{\;2}+\Psi\!I_{0}^{\:2}}\,.
\end{equation} 
This derivative depends on the reference point only through an additive constant.

The motion equations are obtained by setting the derivatives, $\partial{S}/\partial{\alpha_{n}^{}}$, equal to constants, $\beta_{n}^{}$.  It can be seen from equation~\eqref{dS dalpha}, however, that setting $q_{\nu}^{} = (q_{\nu}^{})_{0}^{}$ in $\partial{S}/\partial{\alpha_{n}^{}} = \beta_{n}^{}$ implies that $\beta_{n}^{} = 0$ for each $n$.  Therefore, the motion equations are simply $\partial{S}/\partial{\alpha_{n}^{}} = 0$ or, equivalently,  
\begin{equation*} \dfrac{\Psi\!R\,\dfrac{\partial{\Psi\!I}}{\partial{\alpha_{n}^{}}}-\Psi\!I\,\dfrac{\partial{\Psi\!R}}{\partial{\alpha_{n}^{}}}}{\Psi\!R^{\,2}+\Psi\!I^{2}} = c_{n}^{} = \dfrac{\hslash\left(\!\Psi\!R_{0}^{}\,\dfrac{\partial{\Psi\!I_{0}^{}}}{\partial{\alpha_{n}^{}}}-\Psi\!I_{0}^{}\,\dfrac{\partial{\Psi\!R_{0}^{}}}{\partial{\alpha_{n}^{}}}\right)}{\Psi\!R_{0}^{\;2}+\Psi\!I_{0}^{\:2}}
\end{equation*}
where the $c_{n}^{}$ are new constants that are determined by $\Psi_{0}^{}$ and its derivatives.

For a stationary state of a conservative system, the energy, $E$, may be taken as one of the constants, $\alpha_{n}^{}$, say $\alpha_{\mathscr{N}}^{}$.  Then $S = W(q_{n}^{},(q_{n}^{})_{0}^{},\alpha_{1}^{},\alpha_{2}^{},\ldots,\alpha_{\mathscr{N}-1}^{},E,\gamma)-E\,t$, and the motion equations are
\begin{align*}
	&\dfrac{\partial{W}}{\partial{\alpha_{n}^{}}} = \beta_{n} \qquad \text{for} \qquad n = 1,2,\ldots,\mathscr{N}-1 \\[7pt]
	&\dfrac{\partial{W}}{\partial{E}} = t+\beta_{\mathscr{N}} \;.
\end{align*}
In terms of the function $\psi$, these motion equations become
\begin{align*}	&\dfrac{\psi\!R\,\dfrac{\partial{\psi\!I}}{\partial{\alpha_{n}^{}}}-\psi\!I\,\dfrac{\partial{\psi\!R}}{\partial{\alpha_{n}^{}}}}{\psi\!R^{\,2}+\psi\!I^{2}} = c_{n}^{} \qquad \text{for} \qquad n = 1,2,\ldots,\mathscr{N}-1 \\[7pt]	&\dfrac{\psi\!R\,\dfrac{\partial{\psi\!I}}{\partial{E}}-\psi\!I\,\dfrac{\partial{\psi\!R}}{\partial{E}}}{\psi\!R^{\,2}+\psi\!I^{2}} = t+c_{\mathscr{N}}^{} \;.
\end{align*}
\subsection{Conservative Systems with One Degree of Freedom}
This section shows that conservative systems with one degree of freedom form a special class for which the theory of discrete extension provides simple, closed-form expressions for the action, momentum, and world-line functions.  It also shows that various equations of the theory are generalizations of the corresponding equations in Newtonian theory.
\subsubsection{Dynamic Equations for Conservative Systems with One Degree of Freedom}

If a conservative system is in a stationary state with energy $E$, the wave function has the form $\Psi(x,t,E) = \psi(x,E)\,\exp(-\imgi\,E\,t/\hslash)$ where $\psi(x,E)$, satisfies the time-independent Schr\"odinger equation
\begin{equation}
\dfrac{\partial^{2}\psi}{\partial{x}^{2}}+\dfrac{2m}{\hslash^{2}}\,[\,E-V(x)\,]\,\psi = 0\,.
\label{Schr eqn 1dof}
\end{equation}
A partial derivative is indicated here because the Hamilton--Jacobi formalism also involves derivatives with respect to the ``constant'' $E$.  Since this equation is real, its most general complex solution has the form $\psi = \mathscr{C}1\,\psi1+\mathscr{C}2\,\psi2$ where $\psi1 = \psi1(x,E)$ and $\psi2 = \psi2(x,E)$ are any two linearly independent, real solutions, and $\mathscr{C}1$ and $\mathscr{C}2$ are arbitrary complex constants.  According to equation~\eqref{abbreviated action} the (abbreviated) action, $W$, depends only on the ratio of the real and imaginary parts of $\psi$. Therefore, as far as $W$ is concerned, $\psi$ can be divided by $\mathscr{C}1$ leaving a general solution of the form

\begin{equation}
\psi = \psi1+(\mathscr{A}+\imgi\,\mathscr{B})\,\psi2
\label{wave function 1dof template}
\end{equation}
where $\mathscr{A}$ and $\mathscr{B}$ are two arbitrary, real constants.  Substituting $\psi R = \psi1+\mathscr{A}\psi2$ and $\psi I = \mathscr{B}\psi2$ into equation~\eqref{abbreviated action} gives
\begin{equation}
W = \hslash\left[\atan\!\left(\dfrac{\mathscr{B}\psi2}{\psi1+\mathscr{A}\psi2}\right)-\atan\!\left(\dfrac{\mathscr{B}\psi2_{0}^{}}{\psi1_{0}^{}+\mathscr{A}\psi2_{0}^{}}\right)\right]
\label{action 1dof template}
\end{equation}
where $\psi1_{0}^{} = \psi1(x_{0}^{},E)$, $\psi2_{0}^{} = \psi2(x_{0}^{},E)$, and $x_{0}^{}$ is the reference point coordinate.

The momentum and its $x$-derivative are then
\begin{equation}
p = \dfrac{\partial{W}}{\partial{x}} = \hslash\left[\dfrac{\mathscr{B}\,[\,\psi1\,\psi2_{x}-\psi2\,\psi1_{x}\,]}{[\,\psi1+\mathscr{A}\psi2\,]^{2}+[\,\mathscr{B}\psi2\,]^{2}}\right] \qquad \text{and}
\label{momentum 1dof template}
\end{equation}
\begin{equation*}
p_{x}^{} = \dfrac{\partial{p}}{\partial{x}} = \hslash\!\left[\dfrac{2\,\mathscr{B}\,(\psi2\,\psi1_{x}-\psi1\,\psi2_{x})\left[\,(\psi1+\mathscr{A}\psi2)(\psi1_{x}+\mathscr{A}\psi2_{x})+\mathscr{B}^{2}\psi2\,\psi2_{x}\,\right]}{\left[\,(\psi1+\mathscr{A}\psi2)^{2}+(\mathscr{B}\psi2)^{2}\,\right]^{2}}\right]
\end{equation*}
where $\psi1_{x} = \partial{\!\psi1}/\partial{x}$ and $\psi2_{x} = \partial{\!\psi2}/\partial{x}$, and where Schr\"odinger's equation~\eqref{Schr eqn 1dof} was used to eliminate the second derivatives, $\psi1_{xx}$ and $\psi2_{xx}$, from the $p_{x}^{}$ equation.

The constants $\mathscr{A}$ and $\mathscr{B}$ can now be evaluated in terms of pairs of physical constants such as the momenta, $p_{1}^{} = p(x_{1}^{},E)$ and $p_{2}^{} = p(x_{2}^{},E)$, at two reference locations, or the momentum and its $x$-derivative, $p_{0}^{} = p(x_{0}^{},E)$ and $(p_{x}^{})_{0}^{} = p_{x}^{}(x_{0}^{},E)$, at one reference location.  Choosing the second alternative, solving for $\mathscr{A}$ and $\mathscr{B}$ in terms of $p_{0}^{}$ and $(p_{x}^{})_{0}^{}$, and substituting into equations~\eqref{wave function 1dof template},~\eqref{action 1dof template}, and~\eqref{momentum 1dof template} gives the general~equations
\begin{align}
 & \psi = \mathscr{C}\!\left[\begin{aligned}&\left\{\hslash\,[\,2\,p_{0}^{}\,(\psi2_{x})_{0}+(p_{x}^{})_{0}^{}\,\psi2_{0}\,]-2\,\imgi\,p_{0}^{\,2}\,\psi2_{0}\right\}\psi1 \\
-&\left\{\hslash\,[\,2\,p_{0}^{}\,(\psi1_{x})_{0}+(p_{x}^{})_{0}^{}\,\psi1_{0}\,]-2\,\imgi\,p_{0}^{\,2}\,\psi1_{0}\right\}\psi2\end{aligned}\,\right]
\label{psi 1dof} \\[7pt]
  & \raisebox{2.28ex}{$W = \dfrac{h}{2\,\pi}\;\catan$}\!\!\!\left[\raisebox{2.28ex}{$\dfrac{2\,p_{0}^{\,2}\,(\psi1_{0}\,\psi2-\psi2_{0}\,\psi1)}{\hslash \! \left[\begin{aligned}&[\,(p_{x}^{})_{0}^{}\,\psi2_{0}+2\,p_{0}^{}\,(\psi2_{x})_{0}\,]\,\psi1\\
-\,&[\,(p_{x}^{})_{0}^{}\,\psi1_{0}+2\,p_{0}^{}\,(\psi1_{x})_{0}\,]\,\psi2\,\end{aligned}\right]}$}\right]
\label{action 1dof} \\[7pt]
  & p = \dfrac{4\,p_{0}^{\,3}\,[\,\psi1_{0}\,(\psi2_{x})_{0}-\psi2_{0}\,(\psi1_{x})_{0}\,]\,(\psi1\,\psi2_{x}-\psi2\,\psi1_{x})}{\left[\begin{aligned}&[\,(p_{x}^{})_{0}^{}\,\psi2_{0}+2\,p_{0}^{}\,(\psi2_{x})_{0}\,]\,\psi1\\
-\,&[\,(p_{x}^{})_{0}^{}\,\psi1_{0}+2\,p_{0}^{}\,(\psi1_{x})_{0}\,]\,\psi2\,\end{aligned}\right]^{2}+\left[\dfrac{2\,p_{0}^{\,2}}{\hslash}(\,\psi2_{0}\,\psi1-\psi1_{0}\,\psi2\,)\,\right]^{2}}\;\;.
\label{momentum 1dof}
\end{align}
\newpage
\noindent The absolute square of equation~\eqref{psi 1dof} is

\begin{equation}
\psi^{*}\psi = \left|\mathscr{C}\right|^{2}\!\!\left[\begin{aligned}&\;\hslash^{2}\!\!\left[\begin{aligned}&[\,(p_{x}^{})_{0}^{}\,\psi2_{0}+2\,p_{0}^{}\,(\psi2_{x})_{0}\,]\,\psi1\\
-\,&[\,(p_{x}^{})_{0}^{}\,\psi1_{0}+2\,p_{0}^{}\,(\psi1_{x})_{0}\,]\,\psi2\,\end{aligned}\right]^{2} \\ +&\left[\,2\,p_{0}^{\,2}\,(\,\psi2_{0}\,\psi1-\psi1_{0}\,\psi2\,)\,\right]^{2}\,\end{aligned}\,\right]\,.
\label{psi-squared 1dof}
\end{equation}

While equation~\eqref{action 1dof} expresses $W$ in terms of solutions of Schr\"odinger's equation, $W$, itself, satisfies the third-order, non-linear equation~\eqref{direct eqn for W}.  Therefore, the constants $p_{0}^{}$ and $(p_{x}^{})_{0}^{}$ together with a trivial additive constant can be viewed as three arbitrary integration constants for this third-order equation.

Two sets of arbitrary constants are discussed in section~\ref{sec:gen H-J eqn}, the order-reduction constants, $\gamma$, and the Hamilton--Jacobi constants, $\alpha_{n}^{}$.  For the conservative systems with one degree of freedom considered here, $p_{0}^{}$ and $(p_{x}^{})_{0}^{}$ comprise the set $\gamma$, and the energy, $E$, comprises the set $\alpha_{n}^{}$.

In equations~\eqref{psi 1dof} and~\eqref{psi-squared 1dof} $\mathscr{C}$ is an arbitrary complex constant that plays no role in the theory.  In equation~\eqref{action 1dof} the function $\catan$ is a continuous, cumulative version of the arctangent function.  Thus, whereas the ordinary arctangent function jumps discontinuously from $\pi/2$ to $-\pi/2$ if its argument jumps from $\infty$ to $-\infty$, the $\catan$ function continues to increase.  To clarify this difference, consider the two functions
\begin{align*}
	&f(x) = \atan[\,a\tan(x)\,] \qquad \qquad \;\;\;\; \text{and} \\[7pt]
	&g(x) = \catan[\,a\tan(x)\,] =  \int_{0}^{x}\left[\dfrac{2\,a}{(1+a^{2})+(1-a^{2})\cos(2\,u)}\right]\!\diffd u\;.
\end{align*}
These functions are identical in the interval $-\pi/2 < x < \pi/2$, and their derivatives of all orders are identical everywhere except at discontinuities.  In general, the function $f(x)$ is discontinuous and periodic with period $\pi$, whereas $g(x)$ is continuous and monotonically increasing. For the special case $a = 1$, however, $f(x) = g(x) = x$.

\begin{figure}
\begin{minipage}{5.7cm}
\includegraphics[scale=.54]{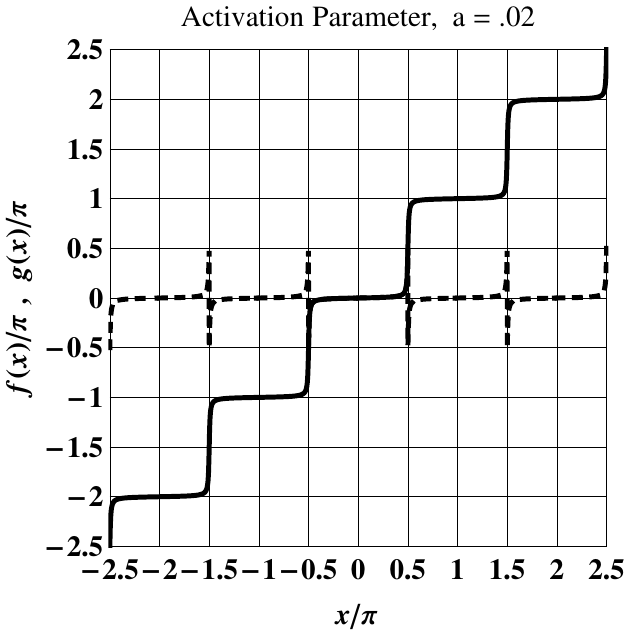} \par \parbox{5.6cm}{\caption{As $a \rightarrow 0$ the function $g(x)$ approaches a staircase function. \label{Staircase}}}
\end{minipage} \quad \;
\begin{minipage}{5.7cm} \vspace{0pt}
\includegraphics[scale=.54]{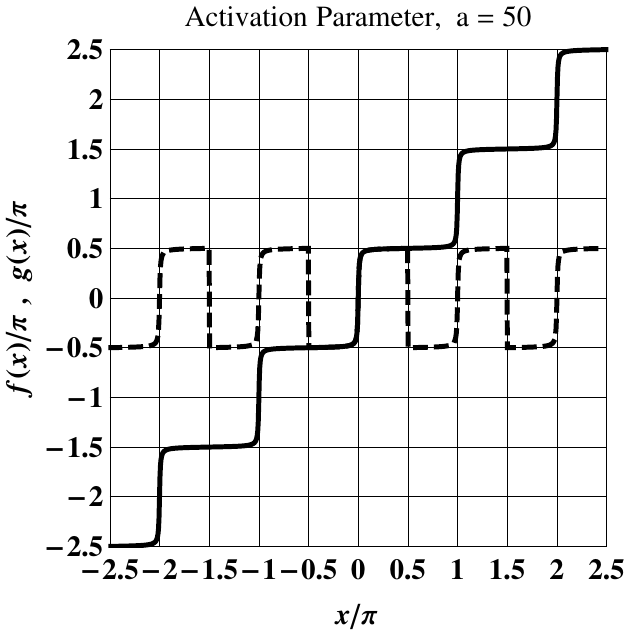} \par \parbox{5.6cm}{\caption{Similarly, as $a \rightarrow \infty$ the function $g(x)$ approaches a shifted staircase function. \label{Shifted Staircase}}}
\end{minipage}
\end{figure}

Figures~\ref{Staircase} and~\ref{Shifted Staircase} illustrate these two functions for $a = .02$ and $a = 50$, respectively.  The dashed and solid lines represent the functions $f(x)$ and $g(x)$, respectively.  Although $g(x)$ is continuous and smooth, it approaches a staircase function as $a \rightarrow 0$.  Similarly, as $a \rightarrow \infty$, it approaches a shifted staircase function.  In the theory of discrete extension the action function for a quantum system is a quasi-discrete function of this type, and a parameter like $a$ characterizes a DEO's level of \emph{quantum activation}, a concept defined below (section~\ref{sec:Quantum Activation}).  Such a parameter will be called an activation parameter.

\newpage

In addition to the action and momentum functions,~\eqref{action 1dof} and~\eqref{momentum 1dof}, the dynamic equations of the theory include the DEO's world-line function.  This function is given by the Hamilton--Jacobi relation $t+\beta_{\mathscr{N}}^{} = \,\partial{W}\!/\partial{E}$.  In taking this derivative it must be observed that the ``constants'' $\psi1_{0}$, $\psi2_{0}$, $(\psi1_{x})_{0}$, $(\psi2_{x})_{0}$, $p_{0}^{}$, and $(p_{x}^{})_{0}^{}$ are all functions of $E$.  Therefore, differentiation of $W$ with respect to $E$ generates the following six new constants:
\begin{align*}
	& \begin{aligned} & (\psi1_{E})_{0}^{} = \left(\dfrac{\partial{\psi1}}{\partial{E}}\right)_{\!0} & \qquad & (\psi1_{xE})_{0}^{} = \left(\dfrac{\partial{\psi1_{x}}}{\partial{E}}\right)_{\!0} \\[7pt]
	& (\psi2_{E})_{0}^{} = \left(\dfrac{\partial{\psi2}}{\partial{E}}\right)_{\!0} && (\psi2_{xE})_{0}^{} = \left(\dfrac{\partial{\psi2_{x}}}{\partial{E}}\right)_{\!0} \end{aligned} \\[7pt]
	& \begin{aligned} & (t_{x}^{})_{0}^{} = \left(\dfrac{\partial{\,t}}{\partial{x}}\right)_{\!0} = \left(\dfrac{\partial^{2}{W}}{\partial{x}\,\partial{E}}\right)_{\!0} = \left(\dfrac{\partial{p}}{\partial{E}}\right)_{\!0} \\[7pt] 
	& (t_{xx}^{})_{0}^{} = \left(\dfrac{\partial^{2}{t}}{\partial{x}^{2}}\right)_{\!0} = \left(\dfrac{\partial^{3}{W}}{\partial{x}^{2}\,\partial{E}}\right)_{\!0} = \left(\dfrac{\partial{p_{x}^{}}}{\partial{E}}\right)_{\!0}\,. \end{aligned}
\end{align*}
The equation of the world-line is then
\begin{equation}
t = \dfrac{2 p_{0}^{}\!\left[\begin{aligned}&2\,p_{0}^{\,2}\!\left[\begin{aligned}
&[\,(\psi2_{x})_{0}\,\psi1_{0}-(\psi1_{x})_{0}\,\psi2_{0}\,]\,[\,\psi2_{E}\,\psi1-\psi1_{E}\,\psi2\,]\\
-\,&[\,(\psi2_{x})_{0}\,\psi1-(\psi1_{x})_{0}\,\psi2\,]\,[\,(\psi2_{E})_{0}\,\psi1-(\psi1_{E})_{0}\,\psi2\,]\\
+\,&[\,\psi2_{0}\,\psi1-\psi1_{0}\,\psi2\,]\,[\,(\psi2_{xE})_{0}\,\psi1-(\psi1_{xE})_{0}\,\psi2\,]
\end{aligned}\right]\\
&\!\!-2\,(t_{x}^{})_{0}^{}\,(\,\psi2_{0}\,\psi1-\psi1_{0}\,\psi2\,)\!\left[\begin{aligned}&[\,(p_{x}^{})_{0}^{}\,\psi2_{0}+p_{0}^{}\,(\psi2_{x})_{0}\,]\,\psi1\\ -\,&[\,(p_{x}^{})_{0}^{}\,\psi1_{0}+p_{0}^{}\,(\psi1_{x})_{0}\,]\,\psi2\,\end{aligned}\right]\,\\
&\!\!+p_{0}^{}\,(t_{xx}^{})_{0}^{}\,(\,\psi2_{0}\,\psi1-\psi1_{0}\,\psi2\,)^{2}
\end{aligned}\right]}{\left[\begin{aligned}&[\,(p_{x}^{})_{0}^{}\,\psi2_{0}+2\,p_{0}^{}\,(\psi2_{x})_{0}\,]\,\psi1\\
-\,&[\,(p_{x}^{})_{0}^{}\,\psi1_{0}+2\,p_{0}^{}\,(\psi1_{x})_{0}\,]\,\psi2\,\end{aligned}\right]^{2}+\left[\,\dfrac{2\,p_{0}^{\,2}}{\hslash}\,(\,\psi2_{0}\,\psi1-\psi1_{0}\,\psi2\,)\,\right]^{2}}
\label{world-line 1dof}
\end{equation}
where the time origin has been chosen so that $\beta_{\mathscr{N}}^{} = 0$.  That is, $t = 0$ when there is a DEO point at the reference point, $x = x_0^{}$.

Note that even though $\psi$ is anti-symmetric on interchange of $\psi1$ and $\psi2$, $\psi^{*}\psi$ and the dynamic variables, $W$, $p$, and $t$, are symmetric as required.  

If $x = \xi$ is the coordinate of a DEO point, then the velocity of that DEO point is $\textsl{v} = 1/\left(\partial{t}/\partial{x}\right)_{x = \xi}^{} = 1/t_x^{}(\xi)$.

The behavior of the corresponding Newtonian system is uniquely determined by two constants: the energy, $E$, and the reference position, $x_{0}^{}$.  In the theory of discrete extension, however, the behavior of the system also depends on the four non-Newtonian constants, $p_{0}^{}$, $(p_{x}^{})_{0}^{}$, $(t_{x}^{})_{0}^{}$, and $(t_{xx}^{})_{0}^{}$, that appear in equation~\eqref{world-line 1dof}.  There are two alternative ways to view these constants.  First, they can be viewed simply as four independent, arbitrary constants.  Second, once a value for $E$ has been specified as part of the Newtonian data, they can be viewed as the values of two arbitrary functions, $p_{0}^{}(E)$ and $(p_{x}^{})_{0}^{}(E)$, and their derivatives with respect to $E$.
\subsubsection{The Generalized Hamilton--Jacobi Equation, Canonical Equations, and Hamiltonian Function}

The momentum equation,~\eqref{momentum 1dof}, has the form $p = G(x,E)$, and the generalized Hamiltonian function, introduced in section~\ref{sec:gen H-J eqn}, is therefore $K(x,p,E) = p-G(x,E) = 0$.
This equation also defines the system's phase space trajectory for any fixed value, $E$, of the energy.

If the equation $K(x,p,E) = 0$ could be solved for $E$, the result would be the Hamiltonian in its explicit, ordinary form, $H = H(x,p) = E$.  As it is, however, the equation $K(x,p,E) = p-G(x,E) = 0$ is an implicit equation for $H$.  Therefore, in terms of the functions $G(x,E)$ and $W\!(x,E)$, Hamilton's canonical equations,
\begin{setlength}{\hfuzz}{2pt}
\begin{align*}
	\begin{aligned} & \qquad \qquad \dfrac{\diffdspc x}{\diffdspc t} = \dfrac{\partial{H}}{\partial{p}} & \qquad & \dfrac{\diffd p}{\diffdspc t} = -\dfrac{\partial{H}}{\partial{x}} \,, \\[-7pt]
\intertext{become} \\[-35pt]
	& \qquad \qquad \dfrac{\diffdspc x}{\diffdspc t} = \dfrac{1}{\left(\dfrac{\partial{G}}{\partial{E}}\right)} = \dfrac{1}{\left(\dfrac{\partial^{2}{W}}{\partial{x}\,\partial{E}}\right)} && \dfrac{\diffd p}{\diffdspc t} = \dfrac{\left(\dfrac{\partial{G}}{\partial{x}}\right)}{\left(\dfrac{\partial{G}}{\partial{E}}\right)} = \dfrac{\left(\dfrac{\partial^{2}{W}}{\partial^{2}{x}}\right)}{\left(\dfrac{\partial^{2}{W}}{\partial{x}\,\partial{E}}\right)} \;. \end{aligned}
\end{align*}
\end{setlength}

Substituting $\partial{W}\!/\partial{x}$ for $p$ in the generalized Hamiltonian function gives the first-order, generalized Hamilton--Jacobi equation for the system,
\begin{equation*}
\dfrac{\partial{W}}{\partial{x}}-G(x,E) = 0\,,
\end{equation*}
in terms of solutions of Schr\"odinger's equation.  In addition to the Hamilton--Jacobi constant, $\alpha = E$, it contains two arbitrary, non-Newtonian constants, $p_0^{}$ and $(p_{x}^{})_{0}^{}$, which comprise the set of constants, $\gamma$, associated with the reduction of the third-order equation~\eqref{direct eqn for W} to a first order equation.  In Newtonian theory the corresponding equation is, of course,
\begin{equation*}
\dfrac{\partial{W_{N}^{}}}{\partial{x}}-\sqrt{2\,m\,(E-V(x))} = 0\,.
\end{equation*}

If the generalized Hamiltonian function, $K(x,p,E)$, could be derived from the physical properties of the system, as in Newtonian theory, then the wave function would not be needed, and $W$ could be obtained from the integral
\begin{equation*}
W = \int{\!p\;\diffd x} = \int{\!G(x,E)\,\diffd x}\,.
\end{equation*}
In the theory of discrete extension, however, that is impossible, and $W$ must be obtained from solutions of Schr\"odinger's equation.

\subsubsection{Generalized Newtonian Energy Equation}

Making the substitution $\diffd W\!/\diffd x = p$ in equation~\eqref{direct eqn for W} and solving for $E$ leads to the energy equation for the theory of discrete extension, presented here in both expanded and compact forms:
\begin{equation} \label{energy x-deriv}
\left. \begin{aligned}
&	E = \dfrac{p^{2}}{2m}\!\left[1+\left(\dfrac{\hslash}{2}\right)^{\!\!2}\!\left(\dfrac{2\,p\,p_{xx}^{}-3\,p_{x}^{\,2}}{p^4}\right)\right]\!+V(x) = T+V \\[7pt]
& E = \dfrac{p^{2}}{2m}\!\left[1+\left(\dfrac{\hslash}{2}\right)^{\!\!2}\!\dfrac{1}{p_{x}^{}}\,\dfrac{\diffd}{\diffd x}\!\!\left(\dfrac{p_{x}^{\,2}}{p^{3}}\right)\right]\!+V(x) = T+V\,.
\end{aligned} \;\right\}
\end{equation}
Converting the $x$-derivatives to $t$-derivatives gives
\begin{equation} \label{energy t-deriv}
\left. \begin{aligned}
&	E = \dfrac{p^{2}}{2m}\!\left[1+\left(\dfrac{\hslash}{2}\right)^{\!\!2}\!\left(\dfrac{2\,p\left(\dot{x}\,\ddot{p}-\dot{p}\,\ddot{x}\right)-3\,\dot{p}^2\,\dot{x}}{p^4\,\dot{x}^3}\right)\right]\!+V(x) = T+V \\[7pt]
& E = \dfrac{p^{2}}{2m}\!\left[1+\left(\dfrac{\hslash}{2}\right)^{\!\!2}\dfrac{1}{\dot{p}}\,\dfrac{\diffd}{\diffdspc t}\!\!\left(\dfrac{\dot{p}^{2}}{p^{3}\,\dot{x}^{2}}\!\right)\right]\!+V(x) = T+V\,.
\end{aligned} \;\right\}
\end{equation}
These equations reduce to the classical energy equation for $\hslash = 0$.  As in Newtonian theory the total energy is the sum of a kinetic part, $T$, and a potential part, $V$.

If $V(x)<\infty$ for all values of $x$, then $T = E-V(x)$ will also be finite everywhere.  This circumstance implies that both $p$ and $\dot{x}$ are never zero since in equation~\eqref{energy t-deriv} they both appear in the denominator of $T$.  (An exception can occur if the motion is such that the numerator of the quantum term is zero.)  Therefore, in general, a DEO can never have zero momentum and can never come to rest, \emph{i.e.}, can never achieve $\dot{x} = 0$.  In standard quantum mechanics a similar conclusion is implied by the uncertainty principle.

It will be seen below that the restriction of $p$ and $\dot{x}$ to non-zero values is illustrated by the behavior of a harmonic oscillator near the turning points of the corresponding Newtonian oscillator.  This behavior is in sharp contrast to the behavior of a Newtonian oscillator for which $p = \dot{x} = 0$ at a turning point.

\subsubsection{Generalized Newtonian Equations of Motion}

In the theory of discrete extension the dynamic behavior of a conservative system with one degree of freedom is given in closed form by equations~\eqref{momentum 1dof} and~\eqref{world-line 1dof}.  These equations are a result of the Hamilton--Jacobi approach to the problem.  Alternatively, the theory can be formulated as a set of differential equations that are a generalization of the Newtonian equations of motion, $p = m\,\dot{x}$ and $F = \dot{p}$.

For a conservative system with one degree of freedom the Newtonian equations can be obtained by differentiating the Newtonian Hamilton--Jacobi equation,
\begin{equation}
\dfrac{1}{2m}\!\left[\dfrac{\partial}{\partial{x}}W_{N}^{}(x,E)\right]^{2}+V(x)-E = 0\,,
\label{classical H-J eqn 1dof}
\end{equation}
with respect to $E$ and with respect to $x$, making the substitutions $\partial{W_{N}^{}}/\partial{x} = p$ and $\partial{W_{N}^{}}/\partial{E} = t$, and converting $x$-derivatives to $t$-derivatives.  Thus, differentiating \eqref{classical H-J eqn 1dof} with respect to $E$ gives the first Newtonian equation, $p = m\,\dot{x}$.  Differentiating \eqref{classical H-J eqn 1dof} with respect to $x$ gives $p\dot{p} = -m\,\dot{x}\:\diffd V\!/\diffd x$.  Since $p = m\,\dot{x}$, however, this equation reduces to the second Newtonian equation, $\dot{p} = -\,\diffd V\!/\diffd x = F$ where $F$ is the force.  These two Newtonian equations comprise a linear system that is first-order in the time derivatives of both $x$ and $p$.  The initial values $x_{0}^{}$ and $p_{0}^{}$ are required to determine a solution.

A similar procedure can be followed in the theory of discrete extension.  There, equation~\eqref{direct eqn for W} is a generalization of the classical Hamilton--Jacobi equation~\eqref{classical H-J eqn 1dof}.  Differentiating it with respect to $E$ and $x$, converting $x$-derivatives to $t$-derivatives, eliminating $E$, and simplifying leads to the following generalization of the Newtonian equations of motion:
\begin{equation} \label{eqns of motion 1dof}
\left. \begin{aligned}
	& p\!\left[1+\left(\dfrac{\hslash}{2}\right)^{\!\!2}\! \left(\dfrac{p\,\dot{x}\left(4\,\dot{p}\,\ddot{x}-\dot{x}\,\ddot{p}\right)+3\left(p^2\,\ddot{x}^2+\dot{p}^2\,\dot{x}^2\right)-p^2\,\dot{x}\,\xdddot}{p^4\,\dot{x}^4}\right)\right]\! = m\,\dot{x} \\[7pt]
	& \dot{p}\!\left[1+\left(\dfrac{\hslash}{2}\right)^{\!\!2}\!\left(\dfrac{p\,\dot{x}\,\pdddot-3\,\ddot{p}\left(p\,\ddot{x}+\dot{p}\,\dot{x}\right)}{m\,p^2 \dot{p}\,\dot{x}^4}\right)\right] = -\dfrac{\diffd V}{\diffd x} = F\,.\end{aligned} \;\right\}
\end{equation}
A more compact form for these equations is
\begin{equation*}
\begin{aligned}
	& p\!\left[1+\left(\dfrac{\hslash}{2}\right)^{\!\!2}\dfrac{1}{2}\; \!\dfrac{\diffd^{\,2}}{\diffdspc t^2}\!\!\left(\dfrac{1}{p^2\,\dot{x}^2}\!\right)\right]\! = m\,\dot{x} \\[7pt]
	& \dot{p}\!\left[1+\left(\dfrac{\hslash}{2}\right)^{\!\!2} \!\dfrac{\:p^2}{m\,\dot{p}}\,\dfrac{\diffd}{\diffdspc t}\!\!\left(\dfrac{\ddot{p}}{p^3\,\dot{x}^3}\!\right)\right]\! = -\dfrac{\diffd V}{\diffd x} = F\,.
\end{aligned}
\end{equation*}

The de~Broglie--Bohm theory postulates the classical relation, $p = m\,\dot{x}$, between momentum and velocity.  It can be seen that in the theory of discrete extension this simple relation does not hold.

Although these equations of motion reduce to Newton's equations for $\hslash = 0$, they can be integrated only in patches if $\hslash\neq0$.  This restriction is due to the generally multi-valued nature of $x(t)$ and $p(t)$ and to the singularities at $p = 0$ and $\dot{x} = 0$.  

The equations of motion~\eqref{eqns of motion 1dof} comprise a non-linear system that is third order in the time derivatives of both $x$ and $p$.  Therefore, a set of six initial values are required to determine a solution.  In addition to the two Newtonian initial values, $x_{0}^{}$ and $p_{0}^{}$, four non-Newtonian initial values, $\dot{x}_{0}^{}$, $\ddot{x}_{0}^{}$, $\dot{p}_{0}^{}$, and $\ddot{p}_{0}^{}$, are also required.  Note that in the Hamilton--Jacobi formulation above, the world-line equation~\eqref{world-line 1dof} also contains a set of six data elements: $E$, $x_{0}^{}$, $p_{0}^{}$, $(p_{x}^{})_{0}^{}$, $(t_{x}^{})_{0}^{}$, and $(t_{xx}^{})_{0}^{}$.  With the help of the energy equation~\eqref{energy x-deriv}, the four non-Newtonian initial values can be expressed in terms of the Hamilton--Jacobi data as follows:
\begin{equation*}
\begin{array}{ll}
\dot{x}_0^{} = \dfrac{1}{(t_{x}^{})_{0}^{}} \qquad \qquad & \ddot{x}_{0}^{} = -\dfrac{(t_{xx}^{})_{0}^{}}{(t_{x}^{})_{0}^{\,3}}\\ \\[4pt] \dot{p}_{0}^{} = \dfrac{(p_{x}^{})_{0}^{}}{(t_{x}^{})_{0}^{}} \qquad \qquad & \ddot{p}_{0}^{} =  \dfrac{\begin{aligned}&\;\,p_{0}^{\,2}\,(t_{x}^{})_{0}^{}\!\left[\,2m\!\left(E-V(x_{0}^{})\right)-p_{0}^{\,2}\right] \\ +&\left(\dfrac{\hslash}{2}\right)^{\!\!2}(p_{x}^{})_{0}^{}\left[\,3\,(p_{x}^{})_{0}^{}\,(t_{x}^{})_{0}^{}-2\,p_{0}^{}\,(t_{xx}^{})_{0}^{}\,\right]\end{aligned}}{2\!\left(\dfrac{\hslash}{2}\right)^{\!\!2}\!p_{0}^{}\,(t_{x}^{})_{0}^{\,3}}\,.
\end{array}
\end{equation*}

\subsubsection{Quasi-Newtonian DEOs}

For a given potential function, $V(x)$, let the reference point, $x = x_{0}^{}$, be chosen so that $V(x_{0}^{}) = 0$.  Then there are special values of the reference point parameters, $p_{0}^{}$, $(p_{x}^{})_{0}^{}$, $(t_{x}^{})_{0}^{}$, and $(t_{xx}^{})_{0}^{}$, for which the DEO's behavior is particularly simple.  These special values are the values of the corresponding Newtonian parameters, $p_{N}^{}$, $(p_{x}^{})_{N}^{}$, $(t_{x}^{})_{N}^{}$, and $(t_{xx}^{})_{N}^{}$ at that point.  That is,
\begin{equation*}
\begin{array}{ll}
p_{0}^{} = p_{N}^{} = \sqrt{2\,m\,E} \qquad \qquad& (p_{x}^{})_{0}^{} = (p_{x}^{})_{N}^{} = \dfrac{\dot{p}}{\dot{x}} = -\sqrt{\dfrac{m}{2\,E}}\;V'(x_{0}^{})\\ \\ (t_{x}^{})_{0}^{} = (t_{x}^{})_{N}^{} = \dfrac{1}{\dot{x}} = \sqrt{\dfrac{m}{2\,E}} \qquad \qquad& (t_{xx}^{})_{0}^{} = (t_{xx}^{})_{N}^{} = -\dfrac{\ddot{x}}{\dot{x}^3} = \sqrt{\dfrac{m}{8\,E^{3}}}\;V'(x_{0}^{})
\end{array}
\end{equation*}
where $V'(x) = \diffd V\!/\diffd x$ and where $m$ and $E$ are the mass and energy of the DEO.
For these special values the DEO's action function, phase-space trajectory, and world-line will resemble those of a Newtonian particle that is subject to the same potential.  A DEO of this type will be called \emph{quasi-Newtonian}.

It will be seen below that a quasi-Newtonian free DEO is an unextended object and that its behavior is identical to the behavior of a Newtonian free particle.  It will also be seen that for the harmonic oscillator potential a quasi-Newtonian DEO is unextended and that its behavior is approximately Newtonian as long as its energy is high and it is not near a turning point.

If a reference point is chosen at which both $V(x_{0}^{}) = 0$ and $V'(x_{0}^{}) = 0$, then $(p_{x}^{})_{N}^{} = 0$, and $(t_{xx}^{})_{N}^{} = 0$ at that point.  Therefore, for a DEO to be quasi-Newtonian in such cases, it is required that $(p_{x}^{})_{0}^{} = 0$ and $(t_{xx}^{})_{0}^{} = 0$. For a free DEO $V = V' = 0$ at all points.  For the harmonic oscillator these conditions occur at the equilibrium point.  For the Coulomb potential they occur at $r = \infty$.  In contrast, there is no point at which $V' = 0$ for the linear potential, $V = c\,x$, of a uniform field.
\section{The Free DEO}
The simplest of all systems is the free DEO with one degree of freedom.  The dynamic equations for that system are derived in this section, and the corresponding DEO behavior is illustrated.  This section also introduces the concept of quantum activation, and presents parameters for quantifying an activation level.

Since a free DEO is not a bound state system, the theory places no restrictions on its allowed energy levels.  For bound-state systems, however, the theory leads to quasi-discrete energy levels, a feature of DEO dynamics that will become apparent later in the analysis of the harmonic oscillator.

Let the mass and energy of the free DEO be $m$ and $E$ respectively, and let the reference point be at $x_{0}^{} = 0$.  The potential function is $V(x) = 0$, and two linearly independent solutions of the time-independent Schr\"odinger equation~\eqref{Schr eqn 1dof} are
\begin{equation*}
\psi1(x,E) = \cos[\,k(E)\,x\,] \qquad \text{and} \qquad \psi2(x,E) = \sin[\,k(E)\,x\,]
\end{equation*}
where $k(E) = \sqrt{2\,m\,E}/\hslash$.  The wavelength associated with these wave functions is $\lambda = 2\,\pi/k$.  Since the reference point coordinate is $x_{0}^{} = 0$, the wave function constants needed for equations~\eqref{psi 1dof} through~\eqref{world-line 1dof} are
\begin{align*}
& \psi1_{0}^{} = 1 & \qquad & (\psi1_{x})_{0}^{} = 0 & \qquad & (\psi1_{E})_{0}^{} = 0 & \qquad & (\psi1_{xE})_{0}^{} = 0 \\[7pt]
& \psi2_{0}^{} = 0 && (\psi2_{x})_{0}^{} = k && (\psi2_{E})_{0}^{} = 0 && (\psi2_{xE})_{0}^{} = \dfrac{m}{\hslash^{2}\,k}\,.
\end{align*}
The following dimensionless constants are defined in terms of the reference point parameters, $p_{0}^{}$, $(p_{x})_{0}^{}$, $(t_{x})_{0}^{}$, and $(t_{xx})_{0}^{}$, and serve to simplify subsequent equations:
\begin{equation*}
A = \dfrac{p_{0}^{}}{\hslash\:k} \quad\;\;\;\; B = \dfrac{(p_{x})_{0}^{}}{2\,kp_{0}^{}} \quad\;\;\;\; C = \dfrac{\hslash^2\,k^2\,(t_{x})_{0}^{}}{m p_{0}^{}}-1 \quad\;\;\;\; D = \dfrac{\hslash^2 k \bigl[p_{0}^{}\,(t_{xx})_{0}^{}-2(p_{x})_{0}^{}\,(t_{x})_{0}^{}\bigr]}{2\,m\,p_{0}^{\,2}}\,.
\end{equation*}
\subsection{The Wave Function and the Action, Momentum, and World-Line Functions} \label{sec:free DEO dynamics}
Substituting these expressions into equations~\eqref{psi 1dof},~\eqref{action 1dof},~\eqref{momentum 1dof}, and~\eqref{world-line 1dof} gives
\begin{equation} \label{free DEO dynamics}
\left. \begin{aligned}
	& \psi = \cos(k\,x)-B\,\sin(k\,x)+\imgi\:A\sin(k\,x) \\[7pt]
	& W = h\!\left[\dfrac{1}{2\,\pi}\,\catan\!\!\left(\!\dfrac{A\sin(k\,x)}{\cos(k\,x)-B\sin(k\,x)}\!\right)\right] \\[7pt]
	& p = \hslash\,k\!\!\left[\dfrac{A}{\left[\,\cos(k\,x)-B\sin(k\,x)\right]^2+\left[\,A\sin(k\,x)\right]^2}\right]  \\[7pt]
	& t = \dfrac{m}{\hslash\,k^2}\!\!\left[\dfrac{A\!\left[\,k\,x+C\sin(k\,x)\cos(k\,x)+D\sin(k\,x)^2\right]}{\left[\,\cos(k\,x)-B\sin(k\,x)\right]^2+\left[\,A\sin(k\,x)\right]^2}\right]\,.
\end{aligned} \;\right\}
\end{equation}
The corresponding Newtonian functions are
\begin{equation} \label{free particle dynamics}
	W_{N}^{} = h\!\left(\!\dfrac{k\,x}{2\,\pi}\!\right) \qquad \qquad	p_{N}^{} = \hslash\,k \qquad \qquad t_{N}^{} =  \dfrac{m}{\hslash\,k^2}(k\,x)\,.
\end{equation}

If the reference point parameters for a free DEO have the Newtonian values,
\begin{equation*}
p_{0}^{} = \sqrt{2\,m\,E} = \hslash\,k \qquad (p_{x})_{0}^{} = 0 \qquad (t_{x})_{0}^{} = \sqrt{m/{2\,E}} = m/{\hslash\,k} \qquad (t_{xx})_{0}^{} = 0\,,
\end{equation*}
then $A = 1$ and $B = C = D = 0$, and the DEO is quasi-Newtonian by definition.  In equations~\eqref{free DEO dynamics} the wave function then becomes $\psi = \exp(\imgi\:k\,x)$, and the expressions for $W$, $p$, and $t$ reduce to the corresponding Newtonian equations~\eqref{free particle dynamics}.  Therefore, a quasi-Newtonian free DEO is unextended, and it behaves in every respect like a Newtonian free particle.
\subsection{Quantum Activation and Activation Parameters} \label{sec:Quantum Activation}
This section introduces the concept of quantum activation in order to quantify the deviation of a free DEO's state from the quasi-Newtonian state.  Since the behaviors of a Newtonian free particle and a quasi-Newtonian free DEO are identical, such a DEO will be said to have a quantum activation level of zero.  Deviation of the value of any of the constants, $A$, $B$, $C$, or $D$, from its Newtonian value corresponds to non-Newtonian behavior of the DEO.  In that case the DEO will be said to possess some level of quantum activation, and it will be useful to find parameters that quantify that level.

Toward that end, the following four new constants are introduced:
\begin{align*}
	\begin{aligned} & \alpha\!1 = \sqrt{1-\left(\!\dfrac{2\,A}{1+A^2+B^{\,2}}\!\right)^{\!\!2}} & \qquad \quad & \phi\!1 = \dfrac{1}{2}\,\atan2\!\left(2\,B,1-A^2-B^{\,2}\right) \\[7pt]
	& \alpha 2 = \dfrac{1}{2}\sqrt{C^{\,2} + D^{\,2}} && \phi\!2 = \dfrac{1}{2}\,\atan2\!\left(C,D\right) \;. \end{aligned}
\end{align*}
In terms of these new parameters the dynamic equations in~\eqref{free DEO dynamics} become:
\begin{align}
  & W = h\!\left[\dfrac{1}{2\,\pi}\,\catan\!\!\left(\!\dfrac{\pm\sqrt{1-\alpha\!1^2}\:\big[\tan(\,k\,x+\phi\!1)-\tan(\phi\!1)\big]}{(1+\alpha\!1)+(1-\alpha\!1)\tan(\phi\!1)\tan(\,k\,x+\phi\!1)}\!\right)\right] \label{free DEO action} \\[7pt]
  & p = \hslash\,k\!\!\left[\dfrac{\pm\sqrt{1-\alpha\!1^2}}{1+\alpha\!1\cos\!\left[\,2\!\left(\,k\,x+\phi\!1\right)\right]}\right] \label{free DEO momentum} \\[7pt]
  & t = \dfrac{m}{\hslash\,k^2}\!\!\left[\dfrac{\pm\sqrt{1-\alpha\!1^2}}{1+\alpha\!1\cos\!\left[\,2\!\left(\,k\,x+\phi\!1\right)\right]}\right]\!\!
\bigg[\,k\,x-\alpha 2\Big[\!\cos\!\left[\,2\!\left(\,k\,x+\phi\!2\right)\right]-\cos\!\left(2\,\phi\!2\right)\!\Big]\bigg]  \label{free DEO world line}
\end{align}
where the $\pm$ sign on the radicals is equal to the sign of $p_{0}^{}$.

The world-line equation~\eqref{free DEO world line} contains the product of two bracketed factors each with its own $\alpha$ and $\phi$ parameters.  The first of these factors has the same periodic form as the momentum equation~\eqref{free DEO momentum}.

The constants $\alpha\!1$ and $\alpha 2$ quantify the DEO's level of activation and will be called the primary and secondary activation parameters, respectively.  The phase angles $\phi\!1$ and $\phi\!2$, however, play no role in determining the activation level.  Equations~\eqref{free DEO action},~\eqref{free DEO momentum}, and~\eqref{free DEO world line} correspond to quasi-Newtonian behavior if and only if $\alpha\!1 = \alpha 2 = 0$.  In that case the DEO will be said to be unactivated.
\subsection{The de~Broglie Relation and Average Momentum}

Oscillations of the wave function, $\psi(x)$, in~\eqref{free DEO dynamics} have the spatial period $\lambda = 2\pi/k$.  Oscillations of $\rho = \psi^{*}(x)\,\psi(x)$, however, have the spatial period $\lambda/2$ as do oscillations of the momentum function, $p(x)$.

Integrating the momentum equation~\eqref{free DEO momentum} over one spatial period produces an average value, $\overline{p} = \hslash\,k$, independent of the values of $p_{0}^{}$ and $(p_{x})_{0}^{}$ and independent of time.  Therefore, the theory of discrete extension leads to the following relation that resembles the de~Broglie relation of standard quantum theory:
\begin{equation*}
\lambda = \dfrac{2\,\pi}{k} = \dfrac{h}{\hslash\,k} = \dfrac{h}{\overline{p}}\,.
\end{equation*}
Note that while this equation relates $\overline{p}$ to the wavelength of $\psi(x)$, the wavelength of $p(x)$ itself is half that value.
\subsection{Illustrations of Free DEO Dynamics}
\begin{figure}
\begin{minipage}{5.7cm}
\includegraphics[scale=.53]{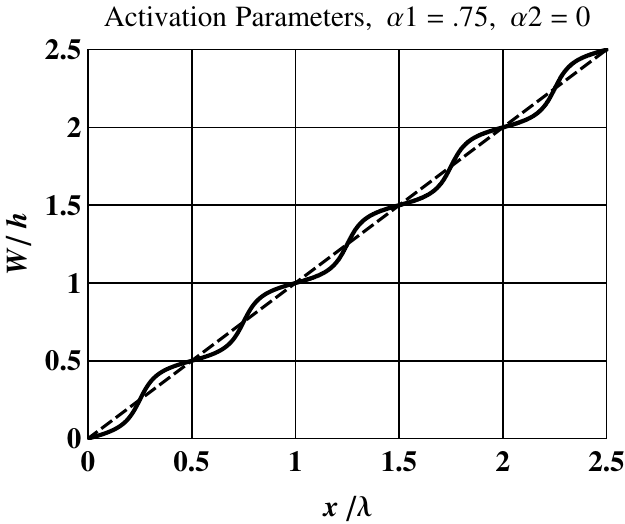} \par \parbox{5.7cm}{\caption{Action function for a moderately activated free DEO.  The dashed line is the action function for the corresponding Newtonian particle.} \label{free DEO action mod activ}}
\end{minipage} \quad \;
\begin{minipage}{5.7cm} \vspace{1pt}
\includegraphics[scale=.53]{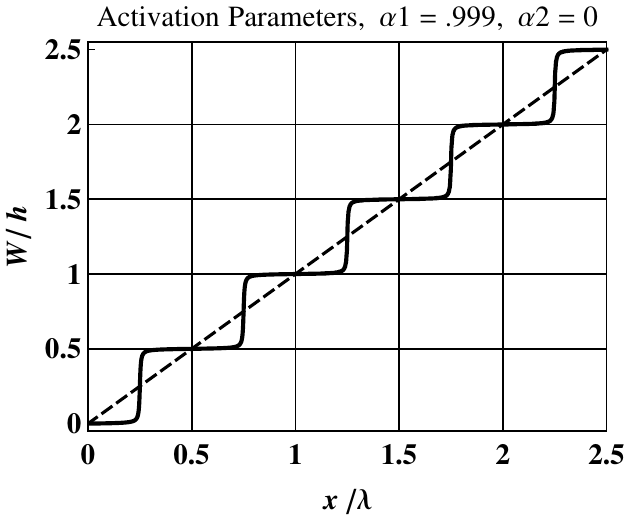} \par \parbox{5.7cm}{\caption{At higher activation levels the action function approaches a staircase function with steps of height $h/2$ and plateaus of width $\lambda\!/2$.} \label{free DEO action high activ}}
\end{minipage} \\[7pt]
\begin{minipage}{5.7cm}
\includegraphics[scale=.53]{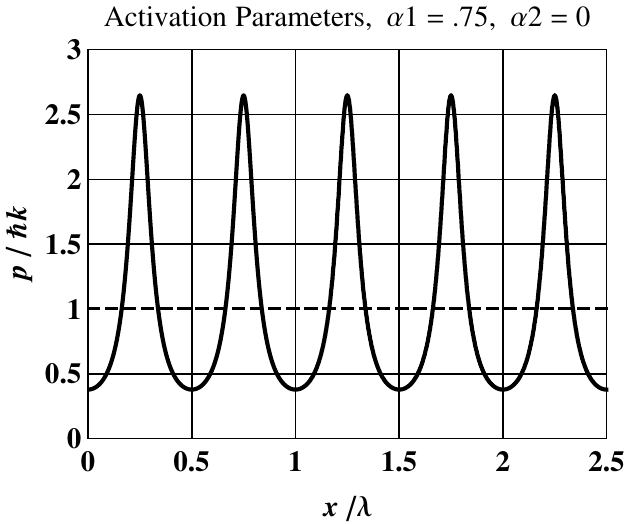} \par \parbox{5.7cm}{\caption{The momentum of an activated free DEO varies periodically with position.  The dashed line shows its average value, $\hslash\,k$.} \label{free DEO momentum mod activ}}
\end{minipage} \quad \;
\begin{minipage}{5.7cm} \vspace{0pt}
\includegraphics[scale=.53]{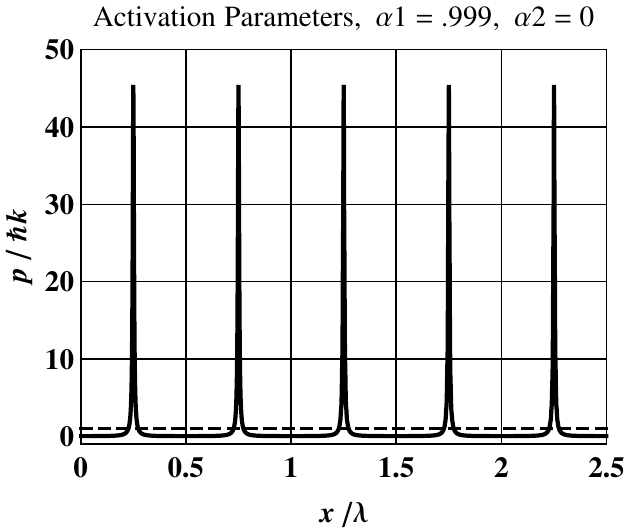} \par \parbox{5.7cm}{\caption{At higher activation levels the momentum function approaches a series of delta functions.  The average value remains $\hslash\,k$.} \label{free DEO momentum high activ}}
\end{minipage} \\[7pt]
\begin{minipage}{5.7cm}
\includegraphics[scale=.53]{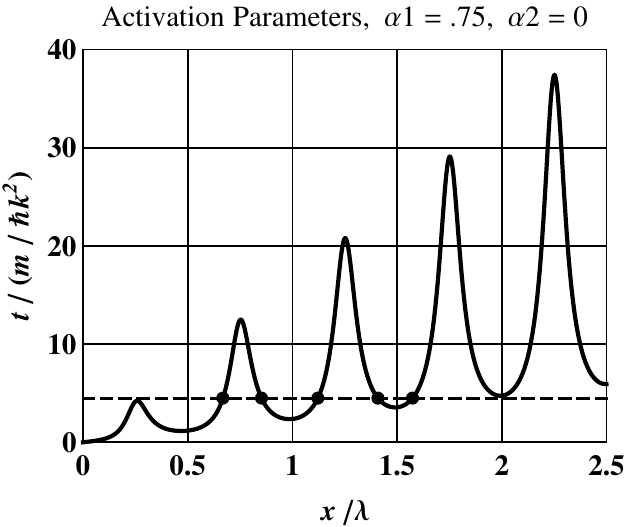} \par \parbox{5.7cm}{\caption{The world-line of an activated DEO is oscillatory.  The number of DEO points remains odd while changing with time in steps of 2.} \label{free DEO world-line mod activ}}
\end{minipage} \quad
\begin{minipage}{5.8cm} \vspace{0pt}
\includegraphics[scale=.545]{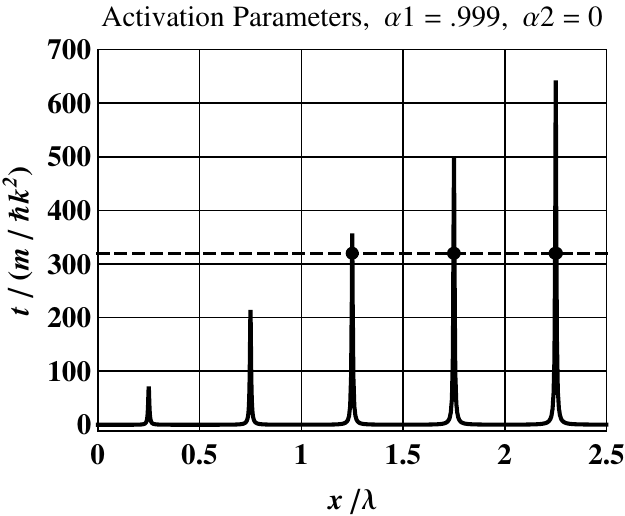} \par \parbox{5.8cm}{\caption{At higher activation levels the world-line approaches a series of delta functions, and the DEO point locations become spatially periodic.} \label{free DEO world-line high activ}}
\end{minipage}
\end{figure}

Figures~\ref{free DEO action mod activ} through~\ref{free DEO world-line high activ} illustrate the behavior of a free DEO for two different levels of quantum activation.  The mass and energy of the DEO are the same in the two cases, and $B = C = D = \phi\!1 = \alpha 2 = 0$ in both cases.  In the first case the DEO is moderately activated with $\,A = .378$, \emph{i.e.}, with $\alpha\!1 = .75$.  In the second case the activation level is significantly higher with $A = .022$, \emph{i.e.}, with $\alpha\!1 = .999$.

In figures~\ref{free DEO action mod activ} and~\ref{free DEO action high activ} the solid curve represents the action function for the activated DEO.  The dashed line represents the action function for the corresponding Newtonian particle or unactivated DEO.  In figure~\ref{free DEO action mod activ} the action function for the moderately activated DEO follows the Newtonian function closely with only minor excursions above and below the Newtonian line. In figure~\ref{free DEO action high activ} the activation level is significantly higher.  In that case the action function still follows the Newtonian line, but it now approaches a staircase function.  Thus, rather than increasing smoothly with position, the action of the highly activated DEO increases in an alternating series of plateaus of width $\lambda\!/2$ and quasi-discrete steps of height $h/2$.

Figures~\ref{free DEO momentum mod activ} and~\ref{free DEO momentum high activ} show that for an activated DEO the momentum is a periodic function of position.  The dashed lines in the figures show the average momentum value, $\hslash\,k$, which is independent of the level of quantum activation.  This average equals the constant value of the momentum of the corresponding Newtonian particle or unactivated DEO.  When the activation level is high, as in figure~\ref{free DEO momentum high activ} the momentum is nearly zero except at a set of quasi-discrete, spatially periodic positions.

Figure~\ref{free DEO world-line mod activ} shows the world-line for the moderately activated DEO.  The dots indicate the positions of the five DEO points that exist at time $t = 4.5\;m/\hslash\,k^2$.  These points are determined by the intersections of the world-line and the dashed line of simultaneity.  It can be seen that as time advances, the DEO, as a whole, moves to the right as new DEO points are created on the right and old DEO points are annihilated on the left.  It is also apparent that the number of DEO points increases or decreases in steps of two and that their total number is always odd.  Figure~\ref{free DEO world-line high activ} shows the world-line for the more highly activated DEO along with the first few DEO points that exist at time $t = 320\;m/\hslash\,k^2$.  The positions of the DEO points are distributed in a nearly periodic pattern.

It can be seen in figures~\ref{free DEO world-line mod activ} and~\ref{free DEO world-line high activ} that, at any given moment, a DEO's various DEO points are distributed over a range of different spatial positions.  Therefore, in general, the action and momentum values associated with distinct DEO points will be unequal.
\subsection{The Fully Activated Free DEO} \label{sec:fully activated free DEO}
It is easily shown that $\alpha\!1$ is real for all values of $A$ and $B$ or, equivalently, for all values of $p_{0}^{}$ and $(p_{x})_{0}^{}$.  Its range is $0 \leq \alpha\!1 \leq 1$.  If $\alpha\!1 = 1$ the DEO will be said to be fully activated.  In that case the bracketed factor common to equations~\eqref{free DEO momentum} and~\eqref{free DEO world line} reduces to the sum of an infinite number of equally-spaced Dirac delta functions:
\begin{equation*}
\lim_{\alpha\!1 \rightarrow 1}\!\left[\dfrac{\sqrt{1-\alpha\!1^2}}{1+\alpha\!1\cos\!\left[\,2\!\left(\,k\,x+\phi\!1\right)\right]}\right] = \!\sum_{n \;=\; -\infty}^{\infty}\!\!\!\!\pi\,\delta\!\big[k\,x+\phi\!1-(n+1\!/2)\,\pi\big]\,.
\end{equation*}

Certain aspects of a fully activated free DEO resemble those of a free particle as described in standard quantum mechanics.  Specifically, as $\alpha\!1 \rightarrow 1$ the dynamic activity of the DEO vanishes, and its DEO points become motionless.  Thus, the DEO approaches a stationary state like that of a free particle in the standard theory.  Like a free particle in the standard theory, there are an infinite number of positions on the real line at which a DEO point could be found.  Furthermore, in both theories the probability of finding the object at a particular position is uniform over the set of the object's possible positions.  However, the sets of possible positions for the object are not the same in the two theories.  According to the standard theory the possible positions at which a free particle could be found comprise a continuous set that covers the entire line.  For a fully activated DEO, on the other hand, the possible positions at which a DEO point could be found comprise only a discrete set of equally spaced points.  It can be seen from the world-lines in figures~\ref{free DEO world-line mod activ} and~\ref{free DEO world-line high activ} that as the activation level of a DEO is raised, its properties approach those just described for a fully activated DEO.
\subsection{Two Types of Quantum Activation}
DEO behavior that is associated with activation due to $\alpha\!1 \neq 0$ differs from the behavior that is associated with $\alpha 2 \neq 0$.  This difference can be understood best by considering the two extreme cases ($\alpha\!1 \neq 0$, $\alpha 2 = 0$) and ($\alpha\!1 = 0$, $\alpha 2 \neq 0$).  DEO motion for the first case is illustrated in figures~\ref{free DEO world-line mod activ} and~\ref{free DEO world-line high activ}.  For increasing values of $x$ the heights of successive peaks in these world-lines increase faster than the heights of successive troughs.  Therefore, as time advances, more and more DEO points are generated, and the DEO spreads out spatially to ever greater widths.

\begin{figure}[ht]
\begin{minipage}{5.7cm}
\includegraphics[scale=.54]{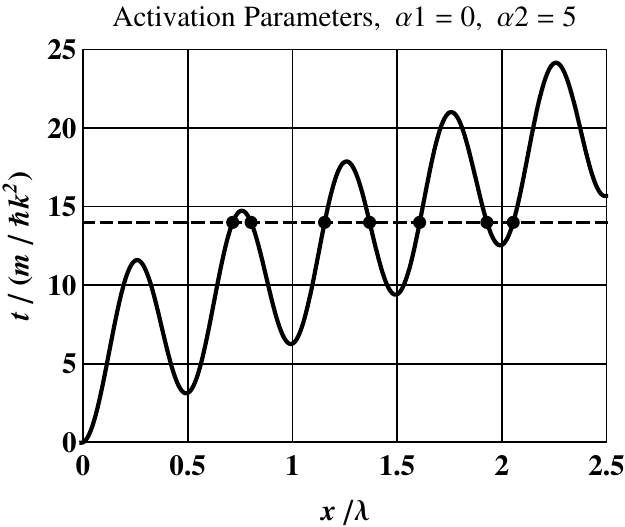} \par \parbox{5.6cm}{\caption{In this example the number of DEO points never exceeds seven.  As the DEO moves, its width never exceeds a value of about $1.3\,\lambda$.} \label{free DEO world-line dlt2 activ}}
\end{minipage} \quad \;
\begin{minipage}{5.7cm} \vspace{0pt}
\includegraphics[scale=.54]{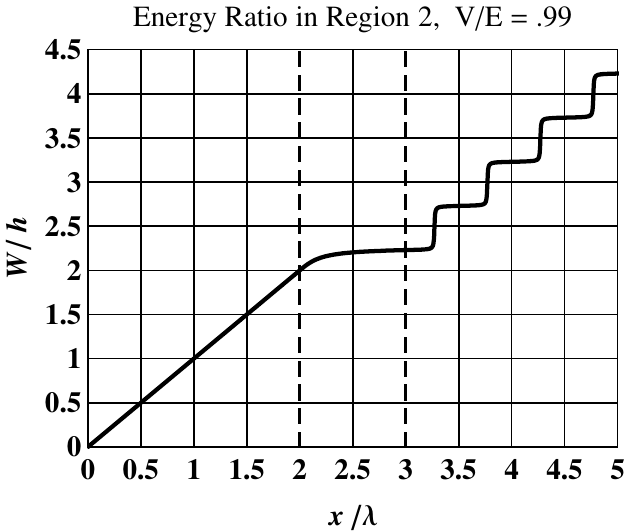} \par \parbox{5.6cm}{\caption{An unactivated free DEO is activated by a barrier encounter.  The resultant primary activation level in region~3 is given by $\alpha\!1 = .998$} \label{barrier action}}
\end{minipage}
\end{figure}

In the second case, illustrated in figure~\ref{free DEO world-line dlt2 activ}, the values of the activation parameters are $\alpha\!1 = 0$ and $\alpha 2 = 5$, and the world-line is a linearly rising sinusoidal curve.  Therefore, as time advances, there is no net change in the number of DEO points.  Furthermore, as the DEO moves, the DEO points do not spread out, but remain within a fixed spatial width, about $1.3\,\lambda$ in this example.  Note that in this second case the action and momentum functions assume their Newtonian forms in~\eqref{free particle dynamics} even though the world-line is non-Newtonian and the DEO is discretely extended.
\section{The Interaction of a Free DEO with a Potential Barrier or Well --- The Emergence of Discrete Extension}
This section analyzes the interaction between a DEO and a rectangular barrier or well.  It is shown that an unactivated DEO becomes activated as a result of the interaction.

A rectangular barrier or well is defined by a potential with a constant value, $V~\!\!\!\neq~\!0$, in the interval between two points, $x = x_{1}^{}$ and $x = x_{2}^{}$ where $x_{1}^{}\!<x_{2}^{}$, and with the value zero elsewhere.  For a barrier $V>0$, and for a well $V\!<0$.  The region $x<x_{1}^{}$ will be referred to as region~1; the region $x_{1}^{} \leq x \leq x_{2}^{}$ will be referred to as region~2; and the region $x_{2}^{}<x$ will be referred to as region~3.

A free DEO with mass $m$ and energy $E>0$ approaches the barrier or well from $x = x_{0}^{}$ in region~1.  In general, this DEO may be activated and may have arbitrary values for the reference point parameters, $p_{0}^{}$, $(p_{x})_{0}^{}$, $(t_{x})_{0}^{}$, and $(t_{xx})_{0}^{}$ (or equivalently $A$, $B$, $C$, and $D$) at $x_{0}^{}$.  The wave functions in the various regions have wave numbers that are functions of energy and are given by
\begin{equation*}
k = \dfrac{\sqrt{2\,m\,E}}{\hslash} \;\; \text{in regions~1 and~3} \qquad \text{and} \qquad \kappa = \dfrac{\sqrt{2\,m\,(V-E)}}{\hslash} \;\; \text{in region~2} \,.
\end{equation*}
The energy ratio in region~2 is $V\!/E = 1+(\kappa/k)^2$.

All of the results of the following analysis are valid whether $E<V$ so that $\kappa$ is real or $V<E$ so that $\kappa$ is imaginary.  Certain functions that are hyperbolic in the first case automatically become trigonometric in the second.  In both cases all action, momentum, and world-line functions are real.

The wave function in the various regions is obtained in the usual way by matching the wave function and its derivative at the boundaries of the regions.  The action function in the various regions is then obtained from this wave function by using equation~\eqref{action 1dof} and requiring the action to be continuous at the boundaries.
\subsection{The Action Function and the Change in Activation Level}
The following dimensionless ``constants'' (actually functions of $E$) serve to simplify the expression for the action function:
\begin{alignat*}{3}
	& a1 = 0 && a2 = \dfrac{\kappa}{k}\! \left\{\begin{aligned}& a1\cos[\,k(x_{1}^{}-x_{0}^{})] \\ +\,& b1\sin[\,k(x_{1}^{}-x_{0}^{})]\end{aligned}\right\} && a3 = \;\:\;\; \left\{\begin{aligned}& a2\cosh[\kappa (x_{2}^{}-x_{1}^{})] \\ +\,& b2\sinh[\kappa (x_{2}^{}-x_{1}^{})]\end{aligned}\right\} \\[4pt]
	& b1 = A && b2 = \;\;\;\;\left\{\begin{aligned}& b1\cos[\,k(x_{1}^{}-x_{0}^{})] \\ -\,& a1\sin[\,k(x_{1}^{}-x_{0}^{})]\end{aligned}\right\} && b3 = \dfrac{\kappa}{k}\! \left\{\begin{aligned}& b2\cosh[\kappa (x_{2}^{}-x_{1}^{})] \\ +\,& a2\sinh[\kappa (x_{2}^{}-x_{1}^{})]\end{aligned}\right\} \\[4pt]
	& c1 = 1 && c2 = \,\dfrac{\kappa}{k}\! \left\{\begin{aligned}& c1\cos[\,k(x_{1}^{}-x_{0}^{})] \\ -\,& d1\sin[\,k(x_{1}^{}-x_{0}^{})]\end{aligned}\right\} && c3 = \;\;\;\;\, \left\{\begin{aligned}& c2\cosh[\kappa (x_{2}^{}-x_{1}^{})] \\ -\,& d2\sinh[\kappa (x_{2}^{}-x_{1}^{})]\end{aligned}\right\} \\[4pt]
	& d1 = B \qquad && d2 = \;\;\;\;\left\{\begin{aligned}& d1\cos[\,k(x_{1}^{}-x_{0}^{})] \\ +\,& c1\sin[\,k(x_{1}^{}-x_{0}^{})]\end{aligned}\right\} \qquad && d3 = \dfrac{\kappa}{k}\! \left\{\begin{aligned}& d2\cosh[\kappa (x_{2}^{}-x_{1}^{})] \\ -\,& c2\sinh[\kappa (x_{2}^{}-x_{1}^{})]\end{aligned}\right\} .
\end{alignat*}
Using these constants the action function is then
\begin{equation*}
W = \left\{\begin{array}{lll} \dfrac{h}{2\,\pi}\,\catan\!\left(\dfrac{a1\cos[\,k(x-x_{0}^{})]+b1\sin[\,k(x-x_{0}^{})]}{c1\cos[\,k(x-x_{0}^{})]-d1\sin[\,k(x-x_{0}^{})]}\right) \quad\; & \text{if} \quad\; x<x_{1}^{} \\ \\ \dfrac{h}{2\,\pi}\, \catan\!\left(\dfrac{a2\cosh[\kappa (x-x_{1}^{})]+b2\sinh[\kappa (x-x_{1}^{})]}{c2\cosh[\kappa (x-x_{1}^{})]-d2\sinh[\kappa (x-x_{1}^{})]}\right) \quad\; & \text{if} \quad\; x_{1}^{} \leq x \leq x_{2}^{} \\ \\ \dfrac{h}{2\,\pi}\, \catan\!\left(\dfrac{a3\cos[\,k(x-x_{2}^{})]+b3\sin[\,k(x-x_{2}^{})]}{c3\cos[\,k(x-x_{2}^{})]-d3\sin[\,k(x-x_{2}^{})]}\right) \quad\; & \text{if} \quad\; x_{2}^{} < x \,.\end{array} \right.
\end{equation*}
In regions~1 and~3 the DEO's primary activation parameter, $\alpha\!1$, is given by
\begin{equation*}
\alpha\!1 = \left\{\begin{array}{lll} \sqrt{1-\left[\dfrac{2\,(a1\,d1+b1\,c1)}{a1^{\,2}+b1^{\,2}+c1^{\,2}+d1^{\,2}}\right]^{\!2}} \qquad & \text{if} \qquad x<x_{1}^{} \\ \\ \sqrt{1-\left[\dfrac{2\,(a3\,d3+b3\,c3)}{a3^{\,2}+b3^{\,2}+c3^{\,2}+d3^{\,2}}\right]^{\!2}} \qquad & \text{if} \qquad x_{2}^{} < x \,.\end{array} \right.
\end{equation*}
Deriving equations for the world-line function and the secondary activation parameter in region~3 are straightforward tasks, but the results are too lengthy to present here.
\subsection{An Example of a Barrier Encounter} \label{sec:barrier interaction}
As an example of an encounter between a DEO and a barrier, consider the case of an unactivated, free DEO approaching the barrier from $x_{0}^{} = 0$ in region~1.  The wavelength of the associated wave function is $\lambda = 2\,\pi/k$, and the boundaries of the barrier are at $x_{1}^{} = 2\,\lambda$ and $x_{2}^{} = 3\,\lambda$.  The energy ratio in region~2 is $V\!/E = .99$.  The action function for this encounter is shown in figure~\ref{barrier action} where the staircase function in region~3 shows that the DEO has been activated by the interaction.  The dashed vertical lines indicate the boundaries of the barrier.

The behavior of the incident DEO is indistinguishable from that of a Newtonian particle.  After its encounter with the barrier, however, the DEO is discretely extended.  The primary activation parameter has changed from $\alpha 1 = 0$ in region~1 to $\alpha\!1 = .998$ in region~3.  There is also a change in the secondary activation parameter from $\alpha 2 = 0$ to $\alpha 2 = 294.9$.

This process of activation through interaction is quite general.  A DEO's activation level almost always increases as the result of an interaction.  It therefore seems evident that any DEO found in nature is likely to be highly activated due to numerous interactions with the environment throughout the DEO's history.  As seen above in section~\ref{sec:fully activated free DEO}, the characteristics of a highly activated free DEO are similar to those of a free particle as described in standard quantum mechanics.  A world filled with naturally occurring DEOs can, therefore, be expected to resemble the world described by standard quantum mechanics.

It is extremely unlikely, but not impossible, for a DEO's activation level to decrease during an interaction.  For the activation level of a given DEO to decrease, the interaction parameters (barrier height, width, and position) would need to be finely tuned to the specific requirements of that DEO.  It is expected that such fine tuning could be accomplished in the laboratory and that a DEO with a low or zero activation level could therefore be artificially generated.  Since objects of that type do not exist in standard quantum mechanics, experimental evidence for them would lend support to the theory of discrete extension.
\section{The Harmonic Oscillator -- An Example of Dynamic, Operator-Free Quantization} \label{sec:Oscillator}

In standard quantum mechanics the harmonic oscillator provides a simple example of a bound-state system with a discrete set of allowed energy levels. In the theory of discrete extension these same energy levels can be determined without recourse to the operator formalism of the standard theory.  The analysis below shows that quantization of an oscillator's action increments and energy levels can be achieved within a mathematical structure similar to that of Newtonian mechanics.  Specifically, it will be seen that the action generated by an oscillator remains nearly constant as the oscillator's energy is increased except at energy values equal to the eigenvalues of the standard theory.  At these special values the action in a full cycle of the oscillator rises abruptly by an amount $h$ from a lower to a higher plateau.  Therefore, judged on the basis of its ability to generate action, the oscillator's behavior is nearly constant for all energies on any given plateau.

Let the mass, spring constant, and energy of the oscillator be $m$, $k$, and $E$ respectively, and let $\omega_{N}^{} = \sqrt{k/m}$ be the angular frequency of the corresponding Newtonian oscillator.  The Newtonian period is $\tau_{N}^{} = 2\pi/\omega_{N}^{}$.  It will be seen that in the theory of discrete extension the angular frequency, $\omega$, of the oscillator depends not only on $k$ and $m$, but also on $E$ and on the DEO's level of quantum activation.

Let the reference point, $x_{0}^{} = 0$, be at the oscillator's equilibrium point.  The system's potential function is then $V(x) = m\,\omega_{N}^{\,2}\,x^{2}/2$, and two linearly independent solutions of Schr\"odinger's equation~\eqref{Schr eqn 1dof} are
\begin{align*}
& \psi1(x,E) = M\!\left[\,\dfrac{1}{4}\!\left(1-\dfrac{2\,E}{\hslash\,\omega_{N}^{}}\right),\,\dfrac{1}{2}\:,\,\dfrac{m\,\omega_{N}^{}}{\hslash}\,x^{2}\,\right]\exp\!\left(-\dfrac{m\,\omega_{N}^{}}{2\,\hslash}\,x^{2}\right) \quad \text{and} \\[9pt]
& \psi2(x,E) = \sqrt{\dfrac{m\,\omega_{N}^{}}{\hslash}}\,x\:M\!\!\left[\,\dfrac{1}{4}\!\left(3-\dfrac{2\,E}{\hslash\,\omega_{N}^{}}\right),\,\dfrac{3}{2}\:,\,\dfrac{m\,\omega_{N}^{}}{\hslash}\,x^{2}\,\right]\exp\!\left(-\dfrac{m\,\omega_{N}^{}}{2\,\hslash}\,x^{2}\right) \quad \text{where} \\[9pt]
& M(a,b,z) = \sum\limits_{n = 0}^{\infty} \dfrac{\Gamma(b)}{\Gamma(a)} \dfrac{\Gamma(a+n)}{\Gamma(b+n)}\dfrac{z^{n}}{n!} \;\: \text{is Kummer's confluent hypergeometric function.}
\end{align*}
The derivatives of $M(a,b,z)$ with respect to $z$ and $a$ are
\begin{align*}
& M_{z}(a,b,z) = \dfrac{\partial M}{\partial z} = \dfrac{a}{b}\,M(a+1,b+1,z)\;\text{,} \\[7pt]
& M_{a}(a,b,z) = \dfrac{\partial M}{\partial a} =  \sum\limits_{n = 0}^{\infty} \bigl[\,\mathrm{Psi}\:\!(a+n)-\mathrm{Psi}\:\!(a)\bigr] \dfrac{\Gamma(b)}{\Gamma(a)} \dfrac{\Gamma(a+n)}{\Gamma(b+n)}\dfrac{z^{n}}{n!}\;\text{,} \quad  \text{and}\\[7pt]
& M_{za}(a,b,z) = \dfrac{\partial^{\:\!2}\!M}{\partial z\;\!\partial a} = \dfrac{a}{b}\,M_{a}(a+1,b+1,z)+\dfrac{1}{b}\,M(a+1,b+1,z)
\end{align*}
where $\mathrm{Psi}\:\!(z)$ is the digamma function.  The values of these functions at $z = 0$ are
\begin{equation*}
M(a,b,0) = 1 \qquad M_{z}(a,b,0) = \dfrac{a}{b} \qquad M_{a}(a,b,0) = 0 \qquad M_{za}(a,b,0) = \dfrac{1}{b}\;.
\end{equation*}
Therefore, at $x_{0}^{} = 0$ the wave function constants for equations~\eqref{psi 1dof} through~\eqref{world-line 1dof} are
\begin{align*}
& \psi1_{0}^{} = 1 & \qquad & (\psi1_{x})_{0}^{} = 0 & \qquad & (\psi1_{E})_{0}^{} = 0 & \qquad & (\psi1_{xE})_{0}^{} = 0 \\[7pt]
& \psi2_{0}^{} = 0 && (\psi2_{x})_{0}^{} = \sqrt{\dfrac{m\,\omega_{N}^{}}{\hslash}} && (\psi2_{E})_{0}^{} = 0 && (\psi2_{xE})_{0}^{} = 0\,.
\end{align*}

Let units of mass, length, and time be defined by $M = m$, $L = \sqrt{\hslash/m\,\omega_{N}^{}\,}$, and $T = 1/\omega_{N}^{}$, respectively.  Let $\mu = 1/L$. The unit of momentum is then $\sqrt{m\,\hslash\,\omega_{N}^{}} = \hslash\,\mu$.  In addition, the following dimensionless constants are defined in terms of the reference point parameters, $p_{0}^{}$, $(p_{x})_{0}^{}$, $(t_{x})_{0}^{}$, and $(t_{xx})_{0}^{}$, and serve to simplify subsequent equations:
\begin{equation*}
A = \dfrac{p_{0}^{}}{\hslash\,\mu} \quad\;\;\;\; B = \dfrac{(p_{x})_{0}^{}}{2\,\mu\! p_{0}^{}} \quad\;\;\;\; C = \dfrac{2\,\hslash\,\omega_{N}^{}\,(t_{x})_{0}^{}}{p_{0}^{}} \quad\;\;\;\; D = \dfrac{\hslash\,\omega_{N}^{} \bigl[p_{0}^{}\,(t_{xx})_{0}^{}-2(p_{x})_{0}^{}\,(t_{x})_{0}^{}\bigr]}{\mu p_{0}^{\,2}}\,.
\end{equation*}

In order to facilitate comparisons with standard quantum mechanics, it will be advantageous to adopt the parameter $\eta = E/\hslash\,\omega_{N}^{}-1/2$ as a dimensionless energy parameter.  It is a continuous version of the energy eigenvalue, $n = 0,1,2,\,\ldots$ , of standard quantum mechanics.  Its range is $-1/2\leq\eta<\infty\,$ in which the negative values correspond to energies in the range $0\leq E<E_{0}^{}$ where $E_{0}^{} = \hslash\,\omega_{N}^{}/2$ is the zero-point energy of the oscillator.

The following abbreviated notation for the hypergeometric functions will also be used to simplify the expressions for the action, momentum, and world-line functions:
\begin{equation*}
Mij(z,\eta) = M\!\left(\dfrac{i-2 \eta-1}{4}\,,\,\dfrac{j}{2}\,,\,z^{2}\right) \;\;\;\;\; \text{and} \;\;\;\;\; Mij_{a}(z,\eta) = M_{a}\!\left(\dfrac{i-2 \eta-1}{4}\,,\,\dfrac{j}{2}\,,\,z^{2}\right).
\end{equation*}
\subsection{Action, Momentum, and World-Line Functions}
Substituting the expressions above into equations~\eqref{action 1dof},~\eqref{momentum 1dof}, and~\eqref{world-line 1dof} leads to the following action, momentum, and world-line functions for the harmonic oscillator:
\begin{align}
	& W = h\!\left[\dfrac{1}{2\,\pi}\,\catan\!\!\left(\dfrac{A \mu x\:M33(\mu x,\eta)}{M11(\mu x,\eta)-B\,\mu x\:M33(\mu x,\eta)}\right)\right] \label{oscillator action} \\[7pt]
	& p = \hslash\,\mu\!\!\left[\dfrac{A\bigl[\,M11(\mu x,\eta)\:M31(\mu x,\eta)+2 \eta \mu^{2}x^{2}\,M33(\mu x,\eta)\,M53(\mu x,\eta)\,\bigr]}{\big[\,M11(\mu x,\eta)-B\,\mu x\:M33(\mu x,\eta)\,\bigr]^{\!2}+\bigl[\,A \mu x\:M33(\mu x,\eta)\,\bigr]^{\!2}}\right] \label{oscillator momentum} \\[7pt]
	& \raisebox{-2ex}{$t = \tau_{N}^{}$}\!\!\left[\raisebox{-2ex}{$\dfrac{\dfrac{A \mu x}{4\,\pi}\!\!\left[\begin{aligned} &  M33(\mu x,\eta)\,M11_{a}(\mu x,\eta)-M11(\mu x,\eta)\,M33_{a}(\mu x,\eta) \\ \! + \; & C\;M11(\mu x,\eta)\,M33(\mu x,\eta)+ D\,\mu x\:M33(\mu x,\eta)^{2}\end{aligned}\,\right]}{\bigl[\,M11(\mu x,\eta)-B\,\mu x\:M33(\mu x,\eta)\,\bigr]^{\!2}+\bigl[\,A \mu x\:M33(\mu x,\eta)\,\bigr]^{\!2}}$}\right]\:\raisebox{-2ex}{.} \label{oscillator world-line}
\end{align}
The corresponding Newtonian functions are
\begin{align}
	& \nonumber W_{N}^{} = h\!\left[\dfrac{1}{4\,\pi}\!\left(\!\!\mu x\sqrt{2 \eta+1-\mu^{2} x^{2}}+(2 \eta+1)\,\asin\!\!\left(\!\frac{\mu x}{\sqrt{2 \eta+1}}\!\right)\!\!\right)\right] \\[7pt]
	& \nonumber p_{N}^{} = \hslash\,\mu\sqrt{2 \eta+1-\mu^{2} x^{2}} \\[7pt]
	& t_{N}^{} = \tau_{N}^{}\!\!\left[\dfrac{1}{2\,\pi}\,\asin\!\!\left(\!\frac{\mu x}{\sqrt{2 \eta+1}}\!\right)\right]\;. \label{Newtonian oscillator world-line}
\end{align}
\subsection{The Quasi-Newtonian Oscillator}
This section illustrates the characteristics and the behavior of a quasi-Newtonian oscillator and thereby identifies certain aspects of DEO dynamics not exhibited by free DEOs.  Even though quasi-Newtonian oscillators are not expected to be found in nature, it is anticipated that they could be created artificially by capturing an unactivated free DEO in an oscillator potential, \emph{i.e.}, by ``turning on'' the oscillator potential at the moment the free DEO reaches the oscillator's equilibrium point.

An oscillator is quasi-Newtonian, by definition, if its reference point parameters have the Newtonian values
\begin{align*}
	& \, p_{0}^{} = \sqrt{2\,m\,E} = \sqrt{m\,\hslash\,\omega_{N}^{}\,(2 \eta+1)} && (p_{x}^{})_{0}^{} = 0 \\[4pt]
	& (t_{x}^{})_{0}^{} = \sqrt{m/2\,E} = \sqrt{m/\hslash\,\omega_{N}^{}\,(2 \eta+1)} && (t_{xx}^{})_{0}^{} = 0 \;.
\end{align*}
The corresponding values of $A$, $B$, $C$, and $D$ are
\begin{equation} \label{quasi-Newtonian parameters}
A = \sqrt{2 \eta+1} \qquad\quad B = 0 \qquad\quad C = 2/2 \eta+1 \qquad\quad D = 0\,.
\end{equation}

\subsubsection{A Comparison of Newtonian and Quasi-Newtonian Dynamics}
\begin{figure}
\begin{minipage}{5.7cm}
\includegraphics[scale=.54]{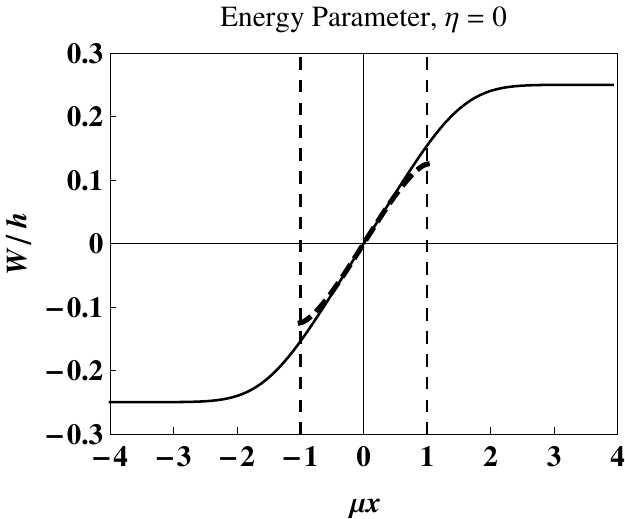} \par \parbox{5.6cm}{\caption{Newtonian and quasi-Newtonian action functions for low energy oscillators.  The dashed vertical lines are at the classical turning points.} \label{quasi-Newt osc action low energy}}
\end{minipage} \quad \;
\begin{minipage}{5.7cm} \vspace{0pt}
\includegraphics[scale=.54]{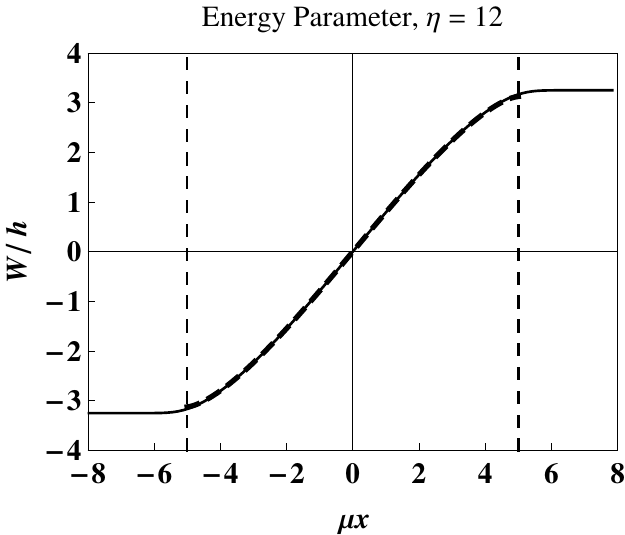} \par \parbox{5.6cm}{\caption{\;\;\; At higher energies the Newtonian and quasi-Newtonian action functions are approximately equal between the classical turning points.} \label{quasi-Newt osc action high energy}}
\end{minipage} \\[6pt]
\begin{minipage}{5.7cm}
\includegraphics[scale=.54]{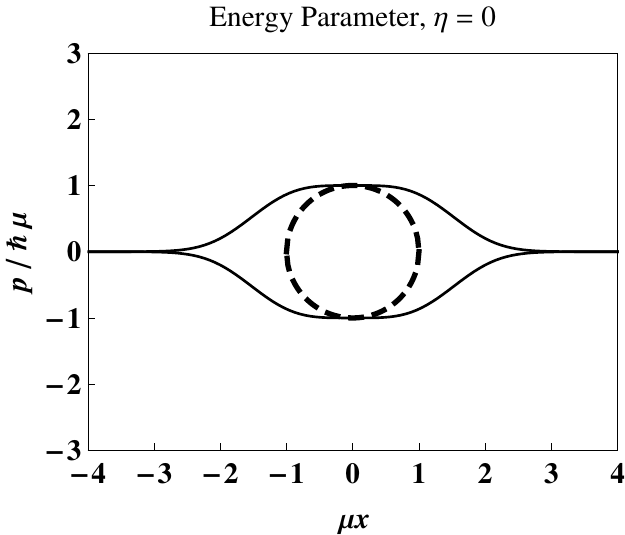} \par \parbox{5.6cm}{\caption{Newtonian and quasi-Newtonian phase space trajectories.  Beyond the classical turning points the DEO escapes momentarily to infinity.} \label{quasi-Newt osc momentum low energy}}
\end{minipage} \quad \;
\begin{minipage}{5.7cm} \vspace{0pt}
\includegraphics[scale=.54]{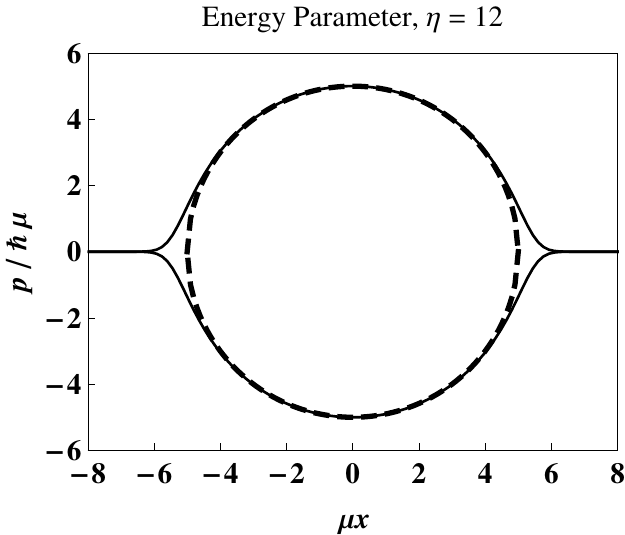} \par \parbox{5.6cm}{\caption{At higher energies the Newtonian and quasi-Newtonian phase space trajectories are approximately equal except near the turning points.} \label{quasi-Newt osc momentum high energy}}
\end{minipage} \\[6pt]
\begin{minipage}{5.7cm}
\includegraphics[scale=.54]{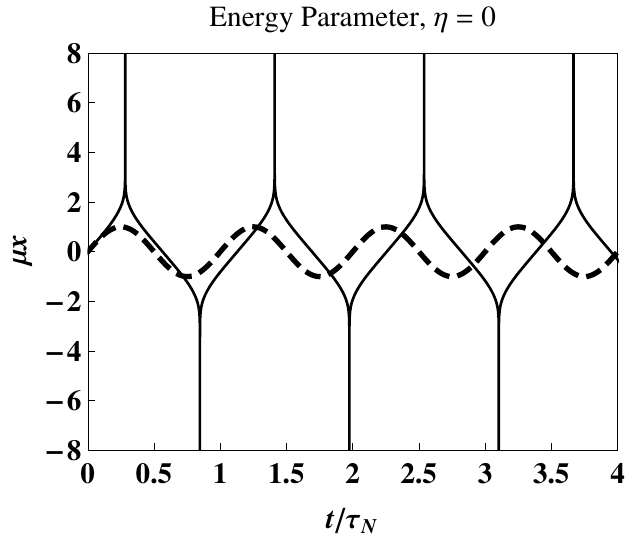} \par \parbox{5.6cm}{\caption{Newtonian and quasi-Newtonian motions.  At low energies the frequencies of the oscillators can differ significantly.} \label{quasi-Newt osc world-line low energy}}
\end{minipage} \quad \,\,
\begin{minipage}{5.7cm} \vspace{0pt}
\includegraphics[scale=.54]{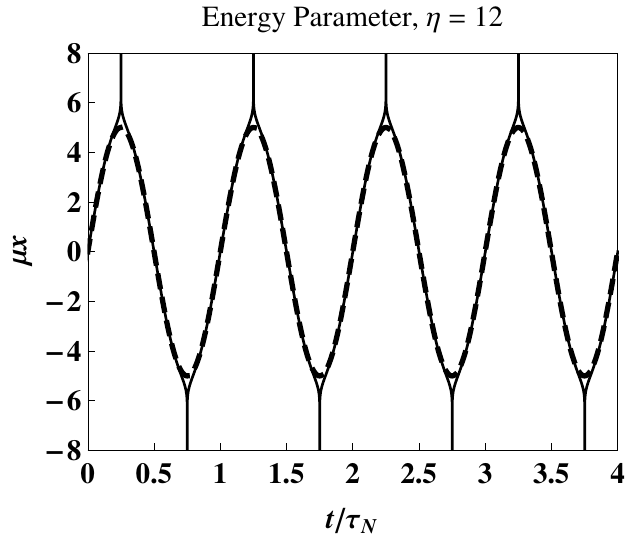} \par \parbox{5.6cm}{\caption{At higher energies the frequencies of Newtonian and quasi-Newtonian oscillators are approximately equal.} \label{quasi-Newt osc world-line high energy}}
\end{minipage}
\end{figure}

Figures~\ref{quasi-Newt osc action low energy} through~\ref{quasi-Newt osc world-line high energy} compare the action, momentum, and motion functions of a quasi-Newtonian oscillator with those of the corresponding Newtonian oscillator.  These comparisons are made for two different energy values, $\eta = 0$ and $\eta = 12$.  In each figure the dashed and solid curves are associated with the Newtonian and quasi-Newtonian oscillators, respectively.  The turning points of the Newtonian oscillator are indicated by vertical dashed lines.

It was seen above in section~\ref{sec:free DEO dynamics} that the behaviors of a Newtonian free particle and a quasi-Newtonian free DEO are identical in all respects.  In contrast, it will be seen here that while Newtonian and quasi-Newtonian oscillators are both unextended objects with periodic motions, their behaviors differ in important ways.  A significant difference concerns the amplitude of the respective oscillations.  While the motion of a Newtonian oscillator is bounded by its turning points, that of a quasi-Newtonian oscillator is unbounded.  In each figure the quasi-Newtonian curve continues beyond the classical turning points and is defined over the range $-\infty<x<\infty$.  The possibility that a quasi-Newtonian oscillator can be found anywhere in this infinite range is consistent with standard quantum mechanics in which the probability density, $\psi^{*}\psi$, for an oscillator's position is non-zero at arbitrarily distant points.

The oscillator action functions for the lower and higher energies are shown in figures~\ref{quasi-Newt osc action low energy} and~\ref{quasi-Newt osc action high energy}, respectively.  In both figures the quasi-Newtonian action function continues to increase for DEO point positions beyond the classical turning points.  For the higher energy case the Newtonian and quasi-Newtonian functions are nearly equal between the turning points.

The oscillator momentum functions (\emph{i.e.}, the phase space trajectories) for the two energies are shown in figures~\ref{quasi-Newt osc momentum low energy} and~\ref{quasi-Newt osc momentum high energy}.  It can be seen that the motions of Newtonian and quasi-Newtonian oscillators are both periodic.  The system point of the Newtonian oscillator moves continuously around its circular trajectory at constant speed.  In contrast, as the momentum of the quasi-Newtonian oscillator approaches zero, its system point accelerates and escapes, momentarily, to spatial infinity before returning for the second half of its cycle.  The momentum is never zero at any finite spatial position, but as it approaches zero, the velocity, $1/\,t_{x}^{}\,$, approaches infinity.

The oscillator motion functions for the two energies are shown in figures~\ref{quasi-Newt osc world-line low energy} and~\ref{quasi-Newt osc world-line high energy}.  They were generated by inverting the world-line equations,~\eqref{oscillator world-line} and~\eqref{Newtonian oscillator world-line}, for the quasi-Newtonian and Newtonian systems, respectively.  The motions of both systems are clearly seen to be periodic, and the periodic escape of the DEO point to infinity is evident.  It is also evident that at low energies the periods of the two oscillators can differ significantly, while at higher energies they are nearly equal in accordance with the correspondence principle.

\newpage

\subsubsection{Probability Density for Position}

The probability density function for the position of a quasi-Newtonian oscillator is derived in this section and is shown to differ from the density function, $\psi^{*}\psi$, specified by the Born rule of standard quantum mechanics.  In principle, this difference provides one of the ways the two theories can be distinguished experimentally.

Quasi-Newtonian and Newtonian oscillators are similar in that both are unextended objects, and both have world-lines that are monotonic between two successive turning points.  As a result of these shared characteristics, the probability density functions for the two oscillators can be derived by similar arguments.  Due to the periodicity and symmetry of an oscillator's motion, its density function can be found by considering only a half-cycle of that motion.

Consider a series of trials in which random times, $\tilde{t}$, are chosen uniformly randomly in the half-period $[\,t\!1\,,\,t2\,]$ where $t\!1$ and $t2$ are the times at which the oscillator arrives at two successive turning points.  The density function for these random times is $P_t^{}(t) = 1/(t2-t\!1)$, a constant.  At each random time the oscillator will be located at some random position, $\tilde{x}$.  The density function for $\tilde{x}$ can be determined from the standard theory of functions of a random variable.  Thus, since $x(t)$ is monotonic in $[\,t\!1\,,\,t2\,]$, the density function for $\tilde{x}$ is simply $P_x^{}(x) = |\,t'(x)|\,P_t^{}(t(x))$ where $t'(x) = \diffdspc t/\diffd x$ is the derivative of the DEO's world-line function~\eqref{oscillator world-line}.  The function $P_t^{}(t(x)) = 1/(t2-t\!1)$ is a constant, independent of $x$.

Since successive turning points of the quasi-Newtonian oscillator are at $x = -\infty$ and $x = \infty$, the half-period, $t2-t\!1$, is equal to $\int_{-\infty}^{\infty}|\,t'(x)|\,\diffd x\;$.  The probability density function for position is therefore
\begin{equation*}
P_x^{}(x) = \dfrac{|\,t'(x)|}{\displaystyle\int_{-\infty}^{\infty}|\,t'(x)|\,\diffd x}\;.
\end{equation*}

For a Newtonian oscillator whose motion is given by $x(t) = X\sin(\omega_{N}^{}\,t)$, a similar argument leads to the density function
\begin{equation*}
(P_{x}^{})_{N}^{}(x) = \left\{ \begin{array}{ll} 1/\pi\sqrt{X^2-x^2} \qquad & \text{if} \qquad -X < x < X \\[5pt] 0 \qquad & \text{otherwise \,.} \end{array} \right.
\end{equation*}

\begin{figure}
\begin{minipage}{5.7cm}
\includegraphics[scale=.535]{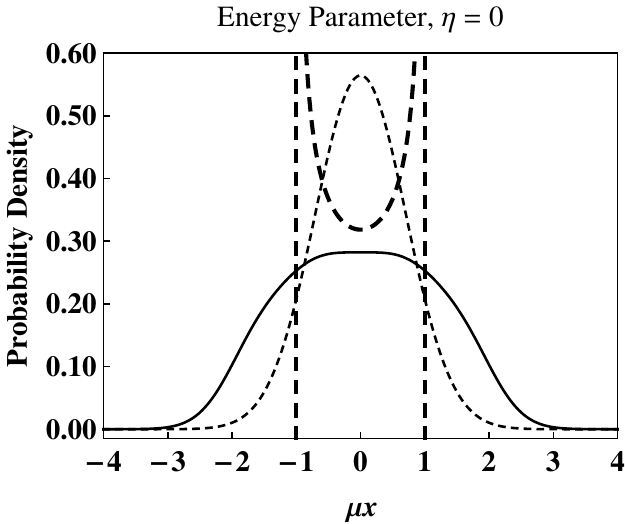} \par \parbox{5.6cm} {\caption{For a quasi-Newtonian oscillator the probability density is non-zero beyond the classical turning points as in standard quantum mechanics.} \label{quasi-Newt osc prob dens low energy}}
\end{minipage} \quad
\begin{minipage}{5.7cm}
\includegraphics[scale=.535]{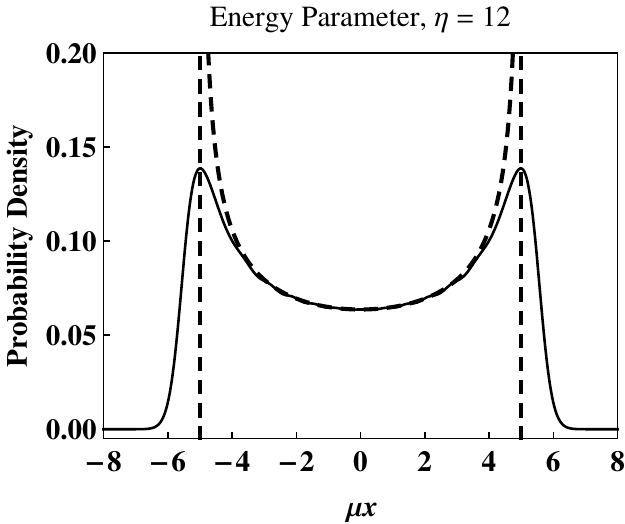} \par \parbox{5.6cm} {\caption{At higher energies the probability density function for a quasi-Newtonian oscillator approximates the Newtonian density function.} \label{quasi-Newt osc prob dens high energy}}
\end{minipage}
\end{figure}

Figures~\ref{quasi-Newt osc prob dens low energy} and~\ref{quasi-Newt osc prob dens high energy} show these probability density functions for two different energies, $\eta = 0$ and $\eta = 12$, respectively.  In both figures the thick, dashed curves and the solid curves are associated with the Newtonian and quasi-Newtonian oscillators, respectively.  The vertical dashed lines are located at the classical turning points.  For a comparison with standard quantum mechanics, figure~\ref{quasi-Newt osc prob dens low energy} also contains a thin, dashed curve which represents the standard density function, $\psi^{*}\psi$, for an energy of $\eta = 0$\,:
\begin{equation*}
\psi^{*}\psi = \sqrt{\dfrac{m\,\omega_{N}^{}}{\hslash\,\pi}}\,\exp{\!\left(\!-\dfrac{m\,\omega_{N}^{}}{\hslash}\,x^{2}\!\right)}\,.
\end{equation*}
There are clear differences in the density functions predicted by the theory of discrete extension and by standard quantum mechanics.

It can be seen in both figures that the probability density for the quasi-Newtonian oscillator is non-zero beyond the Newtonian turning points. Therefore, as in standard quantum mechanics, the oscillator may occasionally be found at arbitrarily large distances from its equilibrium position.

Figure~\ref{quasi-Newt osc prob dens high energy} shows that for higher energies the quasi-Newtonian curve approaches the Newtonian curve in accordance with the correspondence principle.

\subsubsection{Oscillator Frequency and Full-Cycle Action as Functions of Energy}

\begin{figure}[t]
\begin{minipage}{5.7cm} \vspace{0pt}
\includegraphics[scale=.45]{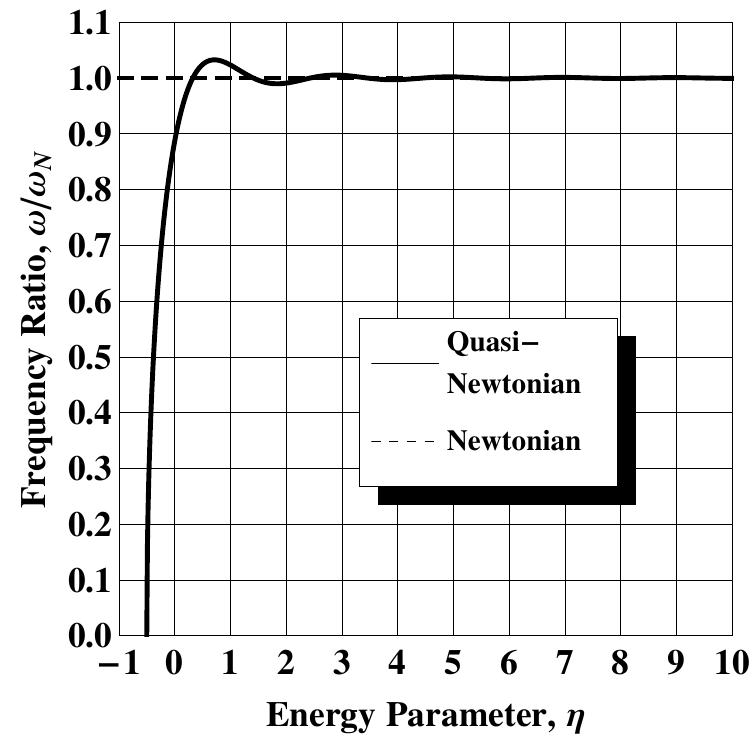} \par \parbox{5.6cm}{\caption{The frequency of a quasi-Newtonian oscillator depends on its energy.  As the energy approaches zero, the oscillation slows to a stop.} \label{quasi-Newt osc freq vs energy}}
\end{minipage} \quad \;
\begin{minipage}{5.7cm}
\includegraphics[scale=.45]{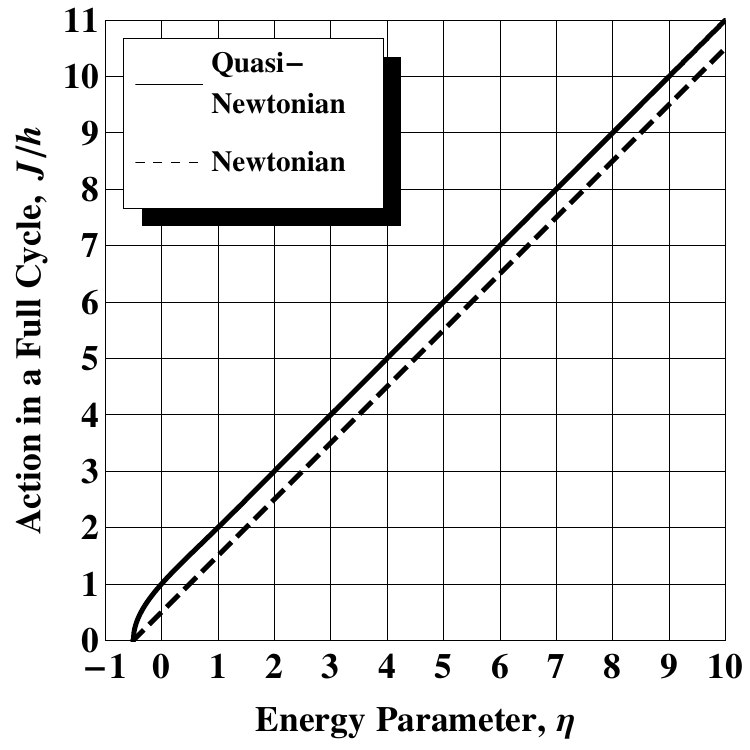} \par \parbox{5.6cm}{\caption{The action generated during a full cycle in units of $h$. As the oscillator's energy approaches zero, the action approaches zero in both systems.} \label{quasi-Newt osc full-cycle action vs energy}}
\end{minipage}
\end{figure}

The motion of a quasi-Newtonian oscillator from $x = 0$ to $x = \infty$ constitutes one quarter of a cycle.  Therefore, using the world-line equation~\eqref{oscillator world-line}, the oscillator's period as a function of its energy is $\tau(\eta) = 4\,t(x = \infty,\eta)$ where the dependence of the world-line function on $\eta$ is shown explicitly.  From the previous section, an alternative expression for the period is $\tau(\eta) = 2\,(t2-t\!1) = 2\!\int_{-\infty}^{\infty}|\,t'(x)|\,\diffd x$.  The angular frequency of a quasi-Newtonian oscillator is then $\omega(\eta) = 2\,\pi/\tau(\eta)$.  In figure~\ref{quasi-Newt osc freq vs energy} the solid curve shows the frequency ratio, $\omega(\eta)/\omega_{N}^{}$, as a function of energy, and the dashed line shows the constant value associated with the Newtonian oscillator.  It can be seen that $\omega(\eta)/\omega_{N}^{} \approx .9$ at the standard zero-point energy, $\eta = 0$.

It can also be seen that the behaviors of Newtonian and quasi-Newtonian oscillators differ markedly at low energies.  If the energy of a Newtonian oscillator is decreased toward zero, its frequency, $\omega_{N}^{}$, remains constant.  In contrast, the frequency of a quasi-Newtonian oscillator is energy-dependent and approaches zero as the oscillator's energy approaches zero, \emph{i.e.}, as $\eta\rightarrow -1/2$.  Therefore, if a quasi-Newtonian oscillator's energy is made smaller and smaller, its period increases without bound as does the time required to make a frequency-based energy measurement.

Figure~\ref{quasi-Newt osc freq vs energy} also shows that, at higher energies, $\omega(\eta)$ approaches the Newtonian frequency in accordance with the correspondence principle.

The action generated in one full cycle of the oscillator is
\begin{equation*}
J(\eta) = \oint{\!p(x,\eta\!)\,\diffd x} \,.
\end{equation*}
Like the oscillator's frequency, the full-cycle action, $J(\eta)$, is a function of energy.  The solid curve in figure~\ref{quasi-Newt osc full-cycle action vs energy} shows this function for a quasi-Newtonian oscillator.  The dashed line shows the corresponding Newtonian function.

In both the Newtonian and quasi-Newtonian cases $J(\eta)$ approaches zero as the oscillator's energy approaches zero.  The Newtonian action approaches zero because the amplitude of the oscillations approaches zero while their frequency remains constant.  In contrast, the quasi-Newtonian action approaches zero because the frequency of the oscillations approaches zero while their amplitude remains infinite.

Figure~\ref{quasi-Newt osc full-cycle action vs energy} also shows that for energies above the standard zero-point energy, $E_{0}^{}$, the quasi-Newtonian action exceeds the Newtonian action by approximately $h/2$.

\subsection{Quantum Mechanical Activation --- A Basis for Energy Quantization}

In this section it is shown that if an oscillator has a high degree of quantum activation, then it is discretely extended, and, furthermore, there are quasi-discrete aspects to its dynamics.  It will be seen, for example, that if the energy of the oscillator is smoothly increased, the full-cycle action, $J(\eta)$, generated by the oscillator remains essentially constant except at integer values of the energy parameter, $\eta$. At each of these special energy values the action makes a quasi-discrete transition to a new plateau.

For the quasi-Newtonian oscillator discussed above, the parameters $A$, $B$, $C$, and $D$ have the Newtonian values~\eqref{quasi-Newtonian parameters}.  To demonstrate the emergence of quantization due to quantum activation, it is sufficient to consider activation due only to the deviation of the parameter $A$ from its Newtonian value.  Thus, let $A = a\,\sqrt{2 \eta+1}$ where the activation parameter, $a$, will be used to raise or lower $A$ from the Newtonian value.  The parameters $B$, $C$, and $D$ will retain their Newtonian values in the following discussion.

\subsubsection{Dynamics of an Activated Oscillator}

\begin{figure}
\begin{minipage}{5.7cm}
\includegraphics[scale=.54]{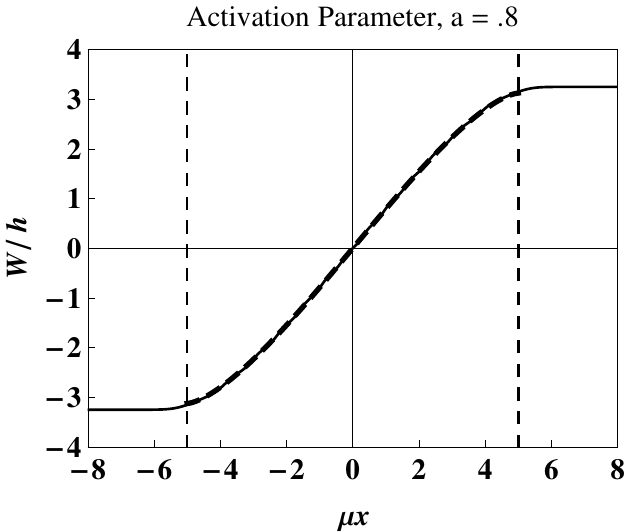} \par \parbox{5.6cm}{\caption{At low activation levels the action function differs very little from the corresponding quasi-Newtonian function of figure~\ref{quasi-Newt osc action high energy}.} \label{osc action low activ}}
\end{minipage} \quad \;
\begin{minipage}{5.7cm} \vspace{0pt}
\includegraphics[scale=.54]{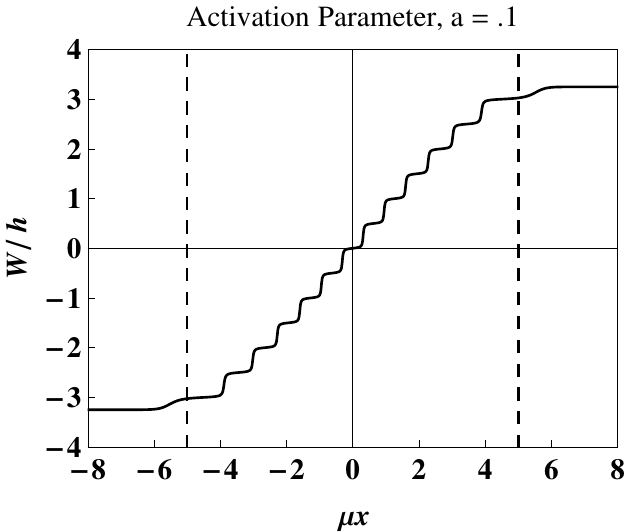} \par \parbox{5.6cm}{\caption{At higher activation levels the action function approaches a staircase function with steps of height $h/2$.} \label{osc action high activ}}
\end{minipage} \\[7pt]
\begin{minipage}{5.7cm}
\includegraphics[scale=.54]{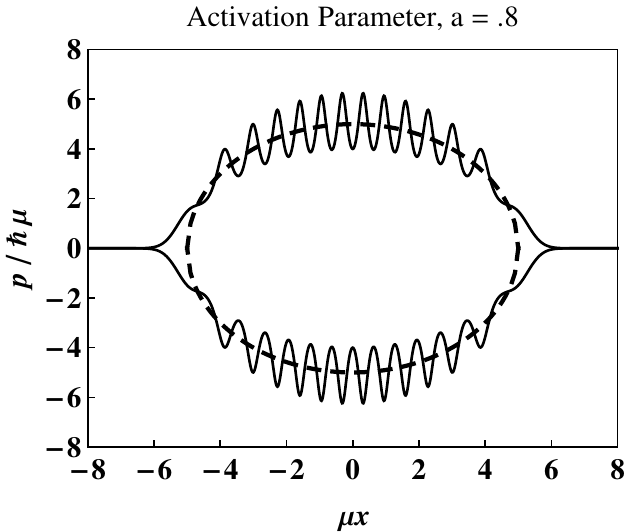} \par \parbox{5.6cm}{\caption{\, The phase space trajectory oscillates around the Newtonian trajectory. The DEO escapes to infinity as in the quasi-Newtonian case.} \label{osc momentum low activ}}
\end{minipage} \quad \!
\begin{minipage}{5.7cm} \vspace{0pt}
\includegraphics[scale=.54]{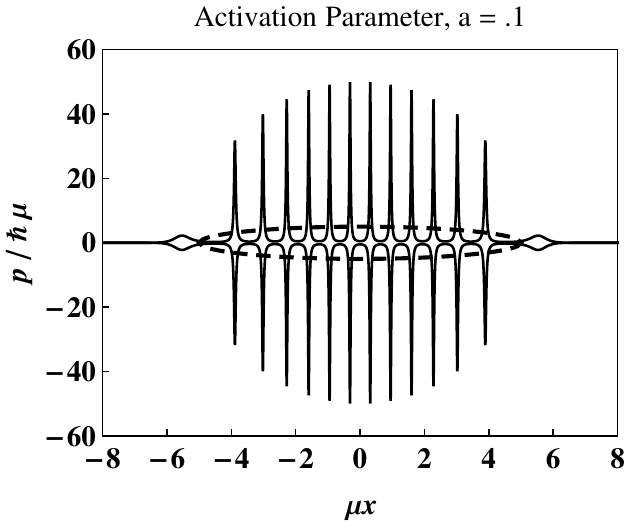} \par \parbox{5.6cm}{\caption{At higher activation levels the oscillations become high and narrow and approach a series of delta functions.} \label{osc momentum high activ}}
\end{minipage} \\[7pt]
\begin{minipage}{5.7cm}
\includegraphics[scale=.54]{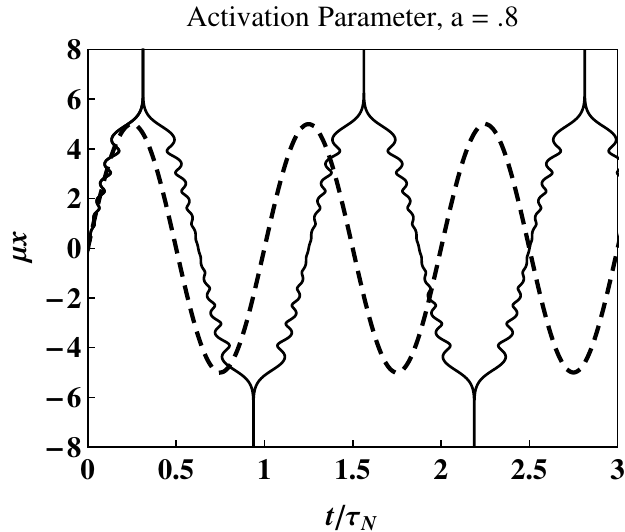} \par \parbox{5.6cm}{\caption{Activation causes the frequency to shift away from the Newtonian value even at high energy levels. Discrete extension begins to emerge.} \label{osc world-line low activ}}
\end{minipage} \quad \;
\begin{minipage}{5.7cm} \vspace{0pt}
\includegraphics[scale=.54]{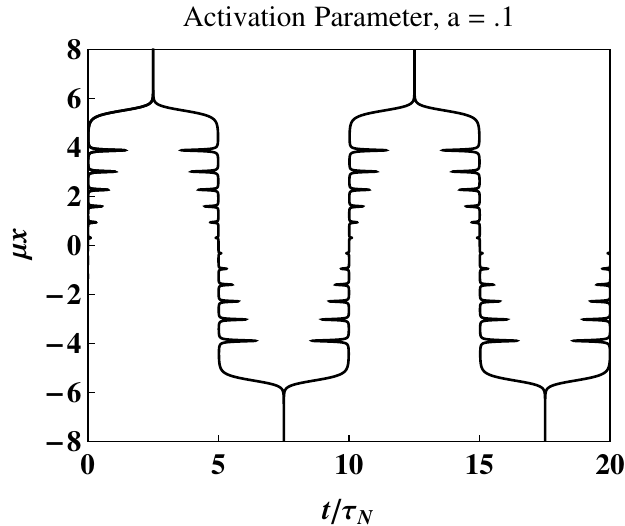} \par \parbox{5.7cm}{\caption{At higher activation levels discrete extension is pronounced. The frequency shift grows, and a stationary state is approached as $a\rightarrow0$.} \label{osc world-line high activ}}
\end{minipage}
\end{figure}

The solid curves in figures~\ref{osc action low activ} through~\ref{osc world-line high activ} illustrate the dynamics of oscillators for both moderate and higher levels of quantum activation.  In both cases $a<1$ so that the value of $A$ is less than the Newtonian value.  For the moderate level $a = .8$, and for the higher level $a = .1$.  In all six figures the energy parameter has the value $\eta = 12$ as in figures~\ref{quasi-Newt osc action high energy},~\ref{quasi-Newt osc momentum high energy}, and~\ref{quasi-Newt osc world-line high energy} for the quasi-Newtonian oscillator.  The dashed curves in the figures show the behavior of the corresponding Newtonian oscillators.  In the interest of visual clarity, however, the Newtonian curves are omitted from figures~\ref{osc action high activ} and~\ref{osc world-line high activ}.  In figure~\ref{osc world-line high activ}, for example, the picture would have been densely overlaid with twenty Newtonian sinusoidal cycles.  As before, the dashed vertical lines represent the classical turning points.

Figure~\ref{osc action low activ} shows that for the moderate activation level the action function differs little from that of the quasi-Newtonian oscillator shown in figure~\ref{quasi-Newt osc action high energy}.  In figure~\ref{osc action high activ}, however, the activation level is higher, and the action function approaches a staircase function, a characteristic of discretely extended systems.

Figures~\ref{osc momentum low activ} and~\ref{osc momentum high activ} show the effect of quantum activation on the oscillator's phase space trajectory.  It can be seen that beyond the turning points the trajectories extend to infinity as in the quasi-Newtonian case shown in figure~\ref{quasi-Newt osc momentum high energy}.  Between the turning points, however, quantum activation causes the trajectories to oscillate around the corresponding Newtonian trajectory.  At higher activation levels as in figure~\ref{osc momentum high activ}, the oscillations become high and narrow and approach a series of delta functions.

The motion functions in figure~\ref{quasi-Newt osc world-line high energy} show that at the higher energy level, $\eta = 12$, the frequencies of Newtonian and quasi-Newtonian oscillators are nearly equal.  In contrast, the motion functions in figures~\ref{osc world-line low activ} and~\ref{osc world-line high activ} show that while quantum activation preserves periodicity, it generates a frequency shift away from the Newtonian frequency even at the higher energy level, $\eta = 12$.  Higher activation levels produce greater frequency shifts, and as $a \rightarrow 0$ the system approaches a stationary state with an infinitely long period.

In both figures~\ref{osc world-line low activ} and~\ref{osc world-line high activ} there exist vertical lines of simultaneity that intersect the DEO motion functions at multiple points.  These intersections demonstrate the existence of discrete extension for both activation levels, but show that it is much more pronounced for the higher activation level.  If $a \rightarrow 0$, the peaks in the world-line function approach a series of delta functions, and the system approaches a stationary state.  In the limit, the height of these peaks and the oscillator's period are both infinite, and, therefore, all vertical lines of simultaneity generate the same static set of DEO points.

Quantum activation also arises if $a>1$ so that the value of $A$ is greater than the Newtonian value.  In the limit as $a \rightarrow \infty$ the peaks in the world-line function again approach a series of delta functions, and the oscillator again approaches a stationary state.  In this case, however, the oscillator's period is infinitesimal.

Whether $a \rightarrow 0$ or $a \rightarrow \infty$, it is evident that the associated probability density for DEO point position is sharply peaked around the delta function locations.

\subsubsection{Quantized Action Increments and Energy Levels} \label{sec:Quantization}

\begin{figure}
\begin{minipage}{5.7cm}
\includegraphics[scale=.45]{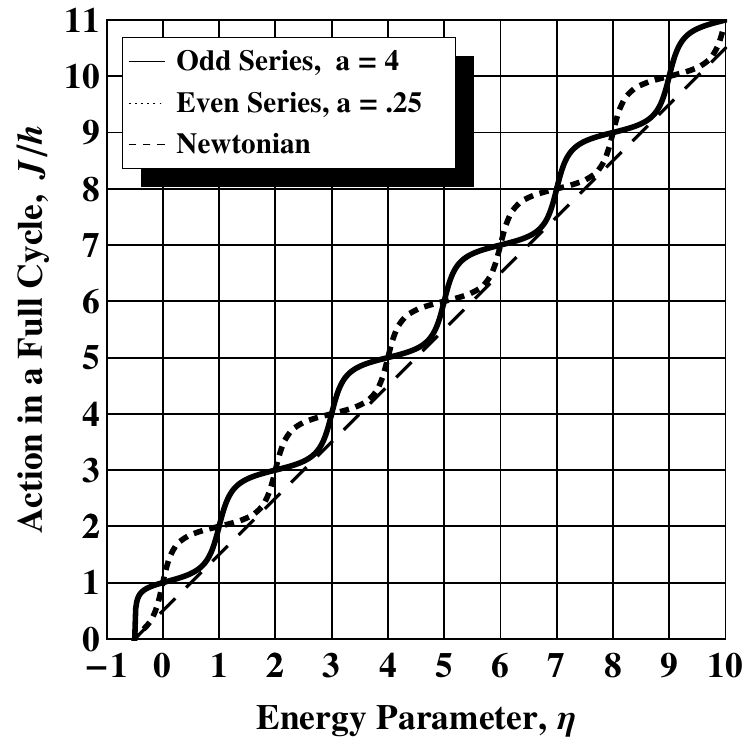} \par \parbox{5.6cm}{\caption{The total action generated in a full cycle in units of $h$. At a moderate activation level the functions begin to take on quasi-discrete forms.} \label{osc full-cycle action vs energy mod activ}}
\end{minipage} \quad \;
\begin{minipage}{5.7cm} \vspace{0pt}
\includegraphics[scale=.45]{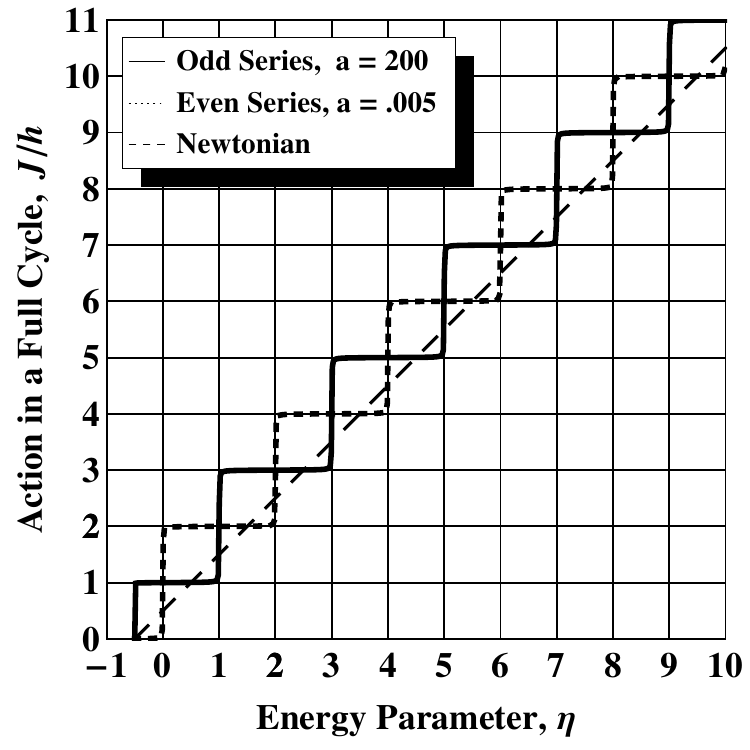} \par \parbox{5.6cm}{\caption{At higher activation levels quasi-discrete action increments of size $h$ occur at the quantized energy values of standard quantum mechanics.} \label{osc full-cycle action vs energy high activ}}
\end{minipage} \\[7pt]
\begin{minipage}{5.7cm}
\includegraphics[scale=.54]{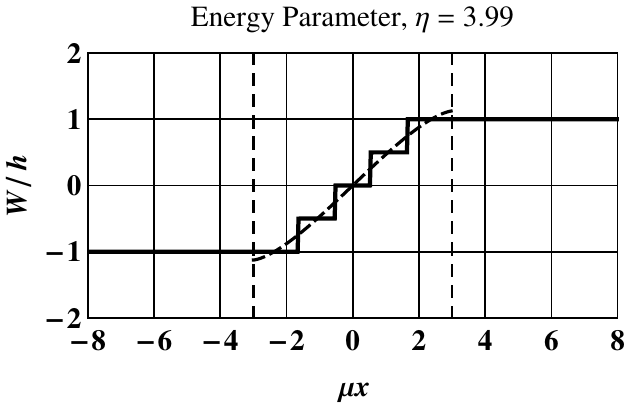} \par \parbox{5.6cm}{\caption{Action as a function of DEO point position within a half cycle.  Quasi-discrete action increments of size $h/2$ occur at four positions.} \label{osc action below trans}}
\end{minipage} \quad
\begin{minipage}{5.7cm} \vspace{0pt}
\includegraphics[scale=.54]{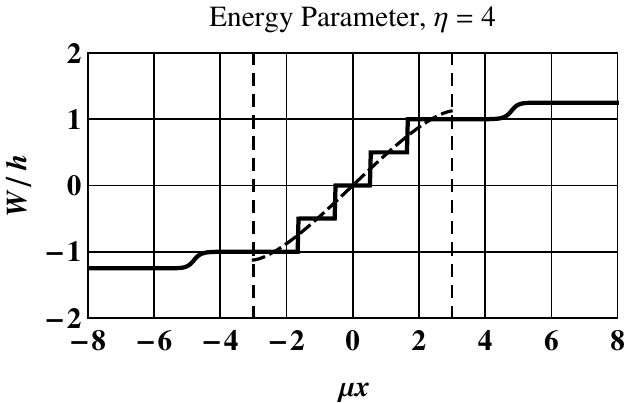} \par \parbox{5.6cm}{\caption{If the energy is increased slightly to the eigenvalue $\eta = 4$, the action function is found to be in the midst of an abrupt transition.} \label{osc action at trans}}
\end{minipage} \\[7pt]
\begin{minipage}{5.7cm}
\includegraphics[scale=.54]{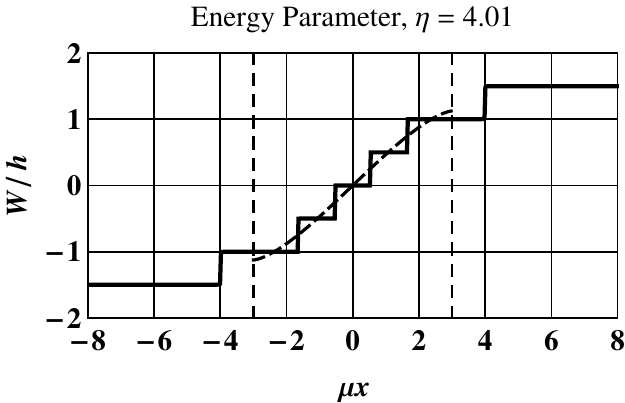} \par \parbox{5.6cm}{\caption{After another small increase in energy the transition is complete, and there are now six action increments of size $h/2$.} \label{osc action above trans}}
\end{minipage} \quad
\begin{minipage}{5.7cm} \vspace{0pt}
\includegraphics[scale=.54]{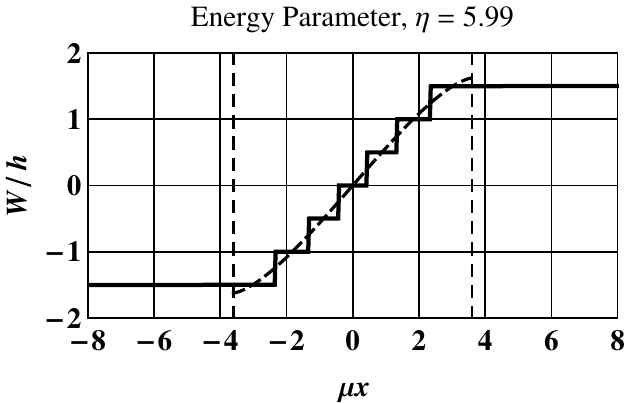} \par \parbox{5.6cm}{\caption{No additional action increments occur as the energy is increased toward a new transition point at the next even eigenvalue, $\eta = 6$.} \label{osc action below next trans}}
\end{minipage}
\end{figure}

Figures~\ref{quasi-Newt osc full-cycle action vs energy},~\ref{osc full-cycle action vs energy mod activ}, and~\ref{osc full-cycle action vs energy high activ} show an oscillator's full-cycle action, $J(\eta)$, as a function of its energy for three different levels of quantum activation.  Figure~\ref{quasi-Newt osc full-cycle action vs energy} is associated with the extreme case of a quasi-Newtonian oscillator.  Figure~\ref{osc full-cycle action vs energy high activ} is associated with the opposite extreme case of a highly activated oscillator.  Figure~\ref{osc full-cycle action vs energy mod activ} corresponds to an oscillator with an intermediate level of activation.  In all three figures the dashed straight line represents the full-cycle action of the corresponding Newtonian oscillator.

Any value of the activation parameter other than $a = 1$ is associated with some level of quantum activation.  The values of $a$ used in figure~\ref{osc full-cycle action vs energy mod activ} for intermediate activation levels are $a = .25$ and $a = 4$.  The values of $a$ used in figure~\ref{osc full-cycle action vs energy high activ} for high activation levels are $a = .005$ and $a = 200$.  It can be seen from these figures that as the oscillator's activation level is increased, the quasi-Newtonian curve of figure~\ref{quasi-Newt osc full-cycle action vs energy} evolves toward a quasi-discrete form in two different ways depending on whether $a \rightarrow 0$ or $a \rightarrow \infty$.

The quasi-discrete curves in figure~\ref{osc full-cycle action vs energy high activ} show that as the energy of a highly activated oscillator is raised, the full-cycle action increases by quasi-discrete increments of magnitude $h$ at successive integer values of the energy parameter, $\eta$.  For values of $\eta$ between successive integers, the action remains on a plateau with a nearly constant value.  These integer values of the continuous variable $\eta$ are, of course, the energy eigenvalues of standard quantum mechanics, and the action increments are those specified in the old quantum theory.  Thus, the Bohr--Sommerfeld quantization rule,
\begin{equation*}
\oint{\!p\;\diffd q} = nh \quad \qquad n = 1,2,3,\ldots\;,
\end{equation*}
is reflected in the theory of discrete extension, not as an ad hoc rule, but as a deduction from the dynamic equations of the theory.

The action increments belong to alternating sequences of even and odd parity.  When $0 < a \ll 1$, quasi-discrete increments of size $2h$ occur at even values of $\eta$, and when $1 \ll a < \infty$ they occur at odd values of $\eta$.

In addition to these action increments at integer values of $\eta$, a special action increment of magnitude $h$ occurs at the unique half-integer value, $\eta = -1/2$.  It is part of the odd-parity sequence and is discussed in the following section in connection with the nature of zero-point energy.

These quasi-discrete action increments can be understood in terms of the behavior of the oscillator's action function~\eqref{oscillator action} as the energy of the system is raised through a transition point.  Figures~\ref{osc action below trans} through~\ref{osc action below next trans} illustrate the evolution of this function around the transition point, $\eta = 4$, in the even-parity sequence.  For this example the oscillator is highly activated with $a = .0001$.  In each of the figures the solid curve shows the oscillator's action as a function of the DEO point coordinate, $x$, where $-\infty < x < \infty$.  The dashed curve is the action function for the corresponding Newtonian oscillator, and the dashed vertical lines represent the classical turning points.

In figure~\ref{osc action below trans} the energy value, $\eta = 3.99$, is just below the transition point.  At this energy level quasi-discrete action increments of size $h/2$ occur at four DEO point positions.  In figure~\ref{osc action at trans} the energy has been raised slightly to the integer value $\eta = 4$.  The action function is now in the midst of an abrupt transition that will lead to the appearance of a new, higher plateau.  In figure~\ref{osc action above trans} the energy has been raised slightly again to the value $\eta = 4.01$, just above the transition point.  The transition is now complete, and action increments of size $h/2$ occur at six DEO point positions.  Taken together, the two new action increments add an additional action amount, $h$, to the half-cycle shown.  Therefore, in the even-parity sequence, the full-cycle action, $J$, increases abruptly by $2h$ as the energy passes through the integer value, $\eta = 4$.

Finally, in figure~\ref{osc action below next trans} the energy has been raised from $\eta = 4.01$ to $\eta = 5.99$.  It can be seen that throughout the energy range $4<\eta<6$ no additional action increments occur in the even-parity sequence.  The oscillator, however, is now on the verge of a new transition at the next even integer, $\eta = 6$.

\subsubsection{The Nature of Zero-Point Energy} \label{sec:Zero-Point Energy}

According to standard quantum mechanics the lowest possible energy of an oscillator is the zero-point energy, $E_{0}^{} = \hslash\,\omega_{N}^{}/2$.  Furthermore, the experimental evidence for the apparent existence of this lower bound is extensive and diverse~\cite{pM-91,kM-01}.  Energies below this value are allowed, however, in the theory of discrete extension, and the theory must explain why oscillators with these lower energies have not been recognized experimentally.  This lack of recognition can be understood if it is assumed that naturally-occurring oscillators have a high degree of quantum activation due to prior interactions with their environments.

Figure~\ref{osc full-cycle action vs energy high activ} from the theory of discrete extension shows the full-cycle action of a highly activated oscillator as a function of the oscillator's energy.  The odd-parity sequence reveals a quasi-discrete action increment of size $h$ near $\eta = -1/2$, \emph{i.e.}, near $E = 0$.  The existence of this action increment leads to the following explanation for the apparent lower bound on energy at $E = E_{0}^{}$.

If the energy of a highly activated oscillator is gradually reduced to $E_{0}^{}$ from a higher value, the full-cycle action decreases in a series of quasi-discrete steps as shown in figure~\ref{osc full-cycle action vs energy high activ}.  If the energy is then decreased further, the action remains nearly constant at the value $h$ and drops to zero suddenly only when $E$ reaches a value very near to zero.  Therefore, judged on the basis of its ability to generate action, the oscillator's behavior at energies in the range $0 < E < E_{0}^{}$ cannot be distinguished from its behavior at $E = E_0^{}$.  For a highly activated oscillator the existence of these lower energies is, therefore, hidden.  Figures~\ref{quasi-Newt osc full-cycle action vs energy} and~\ref{osc full-cycle action vs energy mod activ} show, however, that, for an oscillator with a low or moderate level of quantum activation, the full-cycle action falls to zero gradually as $E \rightarrow 0$.  In those cases, energies below $E_{0}^{}$ may be evident experimentally.

\section{Microscopic Time and Macroscopic Time}
The question arises as to how the behavior of a DEO appears to a macroscopic observer in a Newtonian world.  The theory of discrete extension specifies the behavior of a DEO by providing an equation~\eqref{world-line 1dof} for the DEO's world-line.  The time variable in this equation is specific to the theory of discrete extension and has no \emph{a~priori} connection to the time variable of Newtonian theory.  Furthermore, this world-line inhabits the passive, static realm of space-time and therefore does not, by itself, reveal the active, unfolding dynamics of the DEO.  An active view of the DEO's behavior must be generated by sweeping a line of simultaneity through space-time and noting the progression of its intersections with the DEO's world-line.  To determine the unfolding dynamics as seen by a macroscopic observer, the line of simultaneity must be swept at a rate that represents the uniform flow of Newtonian time.  To accomplish this goal, a relation must be found between the time variable of the theory of discrete extension and the time variable of Newtonian theory.  Such a relation can be determined as follows.

According to the Newtonian law of inertia, equal increments of classical time are associated with equal changes in the position of a Newtonian free particle.  The motion of such a particle can therefore be used to define the uniform flow of Newtonian time.  Similarly, the motion of an unextended (unactivated) free DEO can be used to define the uniform flow of time in the theory of discrete extension.  As seen above in section~\ref{sec:free DEO dynamics}, however, the equations that describe the behaviors of an unextended free DEO and a Newtonian free particle are identical.  Therefore, it will be assumed that Newtonian time and the time defined by the motion of an unextended free DEO are identical.  The observed changes, $\delta x$, in the position of an unextended free DEO thus provide the theory of discrete extension with a self-contained standard for the uniform progression of Newtonian time, namely $\delta t = \delta x\sqrt{m/2E}$.  Therefore, if a line of simultaneity sweeps through space-time according to this standard of uniformity and at this rate, the progression of its intersections with the world-line of any DEO, free or not, will reveal that DEO's behavior as seen by a macroscopic observer.
\section{The Distribution of Mass Among the DEO Points}
At any given moment a DEO will, in general, have a multitude of DEO points.  The question arises as to how the DEO's mass, $m$, is distributed among its various points at that moment.  A hypothesis can be made based on the following considerations:
\begin{enumerate}
		\item At each moment the various DEO points can be divided into two classes based on the sign of the world-line derivative, $\diffdspc t/\diffd x$.  If $\diffdspc t/\diffd x>0$ at the location of a given DEO point, that point will be called a positive point.  If $\diffdspc t/\diffd x<0$, it will be called a negative point. Since all DEO points in a given class have the same character and status, it is natural to assume that they all have the same mass.
	\item As time advances, DEO points may appear or disappear, but they always do so in pairs consisting of one point from each class.
	\item The number of DEO points is always odd with one excess positive point in addition to a number of pairs of positive and negative points.
	\item In a non-relativistic theory the conservation of mass requires that the total mass of all the DEO points remains constant even as their number increases or decreases.
	\item Negative points move in the direction of decreasing $x$, and yet the dynamic equations of the theory generate positive values for the momenta of such points.
\end{enumerate}

Based on these considerations, the theory of discrete extension tentatively assigns a mass value of $m$ to each positive DEO point and $-m$ to each negative DEO point.  The combined mass of each positive/negative pair is then zero, and the mass of the one unpaired, positive point is $m$.  As a result, the total mass of the DEO remains constant in time at the correct value, $m$, even as the number of DEO points rises or falls.  Furthermore, assigning a negative mass value to the negative points is consistent with the positive momentum values generated for such points by the theory.

In order for this assignment of positive and negative mass values to be consistent with the theory's dynamic equations, it is also necessary to assign a positive energy value, $E$, to positive DEO points and a negative value, $-E$, to negative DEO points.  This conclusion can be made plausible by the following considerations:

\begin{enumerate}
	\item At negative DEO points the mass has the negative value $-m$, and the momentum factor, $\sqrt{2\,m\,E}$, would be imaginary if the energy were not also negative.
	\item Consider potential functions, $V(x)$, such as the harmonic oscillator potential, that are proportional to $m$.  For such potentials both terms, $T$ and $V$, on the right side of the energy equation~\eqref{energy x-deriv} are negative when the mass is negative.  Therefore this equation yields a negative energy, $E$, for negative DEO points.
	\item For this same class of potential functions, Schr\"odinger's equation~\eqref{Schr eqn 1dof} and its solutions are invariant with respect to simultaneous sign changes of $m$ and $E$.
\end{enumerate}

Similar considerations with regard to the Coulomb potential and a DEO's electric charge imply that charge should be distributed among the DEO points in the same way that mass is distributed, \emph{i.e.}, with opposite signs at positive and negative DEO points.

It is hoped that the concepts of negative mass and energy introduced here will find a natural interpretation in a relativistic version of the theory of discrete extension.  Just as negative energy electron states in the Dirac electron theory were able to be re-interpreted as positive energy anti-particle states, it is anticipated that a similar re-interpretation for negative DEO points will be possible.  Furthermore, given the equivalence of mass and energy in a relativistic theory, the issues of negative mass and negative energy would be expected to be resolved simultaneously.

In the theory of discrete extension, space is occupied by countless, far-flung pairs of positive and negative DEO points associated with a myriad of highly activated DEOs, free and otherwise.  As illustrated in figure~\ref{free DEO world-line mod activ}, these pairs spontaneously appear and disappear from space as prescribed by the equations of DEO dynamics.  This activity resembles the activity of virtual particle/antiparticle pairs in relativistic quantum field theory.  Therefore, it is conjectured that DEO point pairs and virtual particle/antiparticle pairs would have similar implications regarding the nature of the quantum vacuum.


\begin{thebibliography}{99}

\bibitem{sW-2015}
	Weinberg, S.: Lectures on Quantum Mechanics, Second Edition, p. 102. Cambridge University Press, Cambridge UK (2015)
	
\bibitem{eS-26}
	Schr\"odinger, E.: Quantisation as a Problem of Proper Values (Part I). In: Collected Papers on Wave Mechanics, p. 9. Chelsea, Providence (1982)

\bibitem{dB-51}
	Bohm, D.: Quantum Theory, pp. 609--610. Dover, New York (1989)
	
\bibitem{eS-27}
	Schr\"odinger, E.: The Compton Effect. In: Collected Papers on Wave Mechanics, pp. 124--129. Chelsea, Providence (1982)
	
\bibitem{aL-60}
	Land\'e, A.: From Dualism to Unity in Quantum Physics. Cambridge University Press, London New York (1960)
	
\bibitem{dB-52-1}
	Bohm, D.:
	A Suggested Interpretation of the Quantum Theory in Terms of ``Hidden'' Variables. I and II.
	Phys. Rev. \textbf{85}, 166--193 (1952)
	
\bibitem{dB-52-2}
	\emph{Ibid.}, pp. 170, 173, and 174
	
\bibitem{jB-93}
	Bell, J.: Quantum mechanics for cosmologists. In:
	Speakable and unspeakable in quantum mechanics, p. 128. Cambridge University Press, Cambridge New York (1993)
	
\bibitem{dB-52-3}
	Bohm, D.:
	A Suggested Interpretation of the Quantum Theory in Terms of ``Hidden'' Variables. I.
	Phys. Rev. \textbf{85}, 173 (1952)
		
\bibitem{dB-52-4}
	\emph{Ibid.}, p. 170
	
\bibitem{aM-66}
	Messiah, A.: Quantum Mechanics, pp. 222--224. Wiley, New York (1966)
	
\bibitem{pM-91}
	Milonni, P., Shih, M.:
	Zero-point energy in early quantum theory.
	Am. J. Phys. \textbf{59}, 684--697 (1991)
	
\bibitem{kM-01}
	Milton, K.:
	The Casimir Effect -- Physical Manifestations of Zero-Point Energy. World Scientific, Singapore (2001)
	
\end{thebibliography}
\end{document}